\documentclass[amsmath,twocolumn,superscriptaddress,prx,aps]{revtex4}

\usepackage{epsfig}
\usepackage{graphicx,color}
\usepackage{dcolumn}
\usepackage{bm}
\usepackage{amssymb}
\usepackage{xspace}

\usepackage{soul}

\begin{document}

\setstcolor{red}


\title{Novel magnetic orderings in the kagome Kondo-lattice model}

\author{Kipton Barros}
\affiliation{Theoretical Division and Center for Nonlinear Studies, Los Alamos National Laboratory, Los Alamos, New Mexico 87545, USA}

\author{J\"orn W. F. Venderbos}
\affiliation{IFW Dresden, P. O. Box 27 01 16, D-01171 Dresden, Germany}
\affiliation{Department of Physics, Massachusetts Institute of Technology, Cambridge, Massachusetts 02139, USA}

\author{Gia-Wei Chern}
\affiliation{Theoretical Division and Center for Nonlinear Studies, Los Alamos National Laboratory, Los Alamos, New Mexico 87545, USA}

\author{C. D. Batista}
\affiliation{Theoretical Division and Center for Nonlinear Studies, Los Alamos National Laboratory, Los Alamos, New Mexico 87545, USA}

\date{\today}

\begin{abstract}
We consider the Kondo-lattice model on the kagome lattice and study its weak-coupling instabilities at band filling fractions for which the Fermi surface has singularities. These singularites include Dirac points, quadratic Fermi points in contact with a flat band, and Van Hove saddle points. By combining a controlled analytical approach with large-scale numerical simulations, we demonstrate that the weak-coupling instabilities of the Kondo-lattice model lead to exotic magnetic orderings. In particular, some of these magnetic orderings produce a spontaneous quantum anomalous Hall state.
\end{abstract}

\maketitle

\section{Introduction~\label{introduction}}

In a seminal work, Haldane demonstrated that a magnetic field is not required to induce integer quantum Hall states~\cite{Haldane88}. Adding a complex hopping to the tight-binding  Hamiltonian on the honeycomb lattice opens a gap at the Dirac points of the  electron band structure. This gap leads to a topologically nontrivial electronic state for a half-filled band, i.e. a Chern insulator or quantum anomalous Hall (QAH) state. The key characteristic of this state is the appearance of chiral edge channels at the sample boundaries, in which current can flow along one direction only, implying dissipationless charge transport due to the absence of backscattering. This property makes QAH states attractive for ultra-low-power consumption applications. One of the current challenges in condensed matter physics is to find ways of stabilizing a  QAH state at ambient temperature. 

Many paths toward realization of the QAH effect have been suggested, most of which are limited to graphenelike Dirac systems. Proposals include spin-orbit coupled magnetic semi-conductors~\cite{Qi06}, spin-orbit coupled ad-atoms on graphene~\cite{Zhang12}, and spin-polarized QAH states~\cite{Liu08}. Experimental signatures of QAH have been detected~\cite{Taguchi01,Machida07,Takatsu10} and advanced nanostructures such as ``molecular graphene''~\cite{Gomes12} open the possibility of controlled manipulation of Dirac fermions. A proposal based on doped magnetic topological insulator materials~\cite{Yu10} has led to the first robust observation of the QAH state very recently~\cite{Chang13}. Another proposed scenario involves spontaneous chiral symmetry breaking due to electron-electron interactions at Dirac points~\cite{Raghu08,Weeks10,Grushin13}. However, relatively large electron-electron Coulomb interactions are required to induce chiral states due to the vanishing density of electron states.~\cite{Daghofer14,Martinez13}.

An alternative mechanism for robust, high temperature QAH states has generated much interest~\cite{ohgushi00,martin08,kato10,Akagi10,Kumar10}. In certain correlated multi-orbital compounds, conduction electrons are coupled to localized magnetic moments (spins) at each lattice site. The spin may arise, for example, from $t_{2g}$-electrons in transition-metal oxide materials~\cite{Hopkinson02,chern11} or $f$-electrons in Lanthanide based materials~\cite{Hewson}. QAH states are possible if the spin ordering is {\it noncoplanar}, in which case the local Berry curvature (scalar spin chirality) acts as an  effective magnetic field on the orbital motion of the conduction electrons. This mechanism does not require band structures with Dirac points. QAH states may also arise from weak-coupling instabilities at quadratic band crossings~\cite{sun09,volovik-book,Sun11,Uebelacker11,chern12} and nested Fermi surfaces~\cite{martin08}. Nonzero chirality implies broken time-reversal and spatial parity symmetries, a necessary ingredient for QAH states.  Note that these discrete symmetries can be spontaneously broken at finite temperature in two-dimensions, even though the continuous $SU(2)$ spin rotational symmetry has to remain intact.

\begin{figure}
\includegraphics[width=1.0\columnwidth]{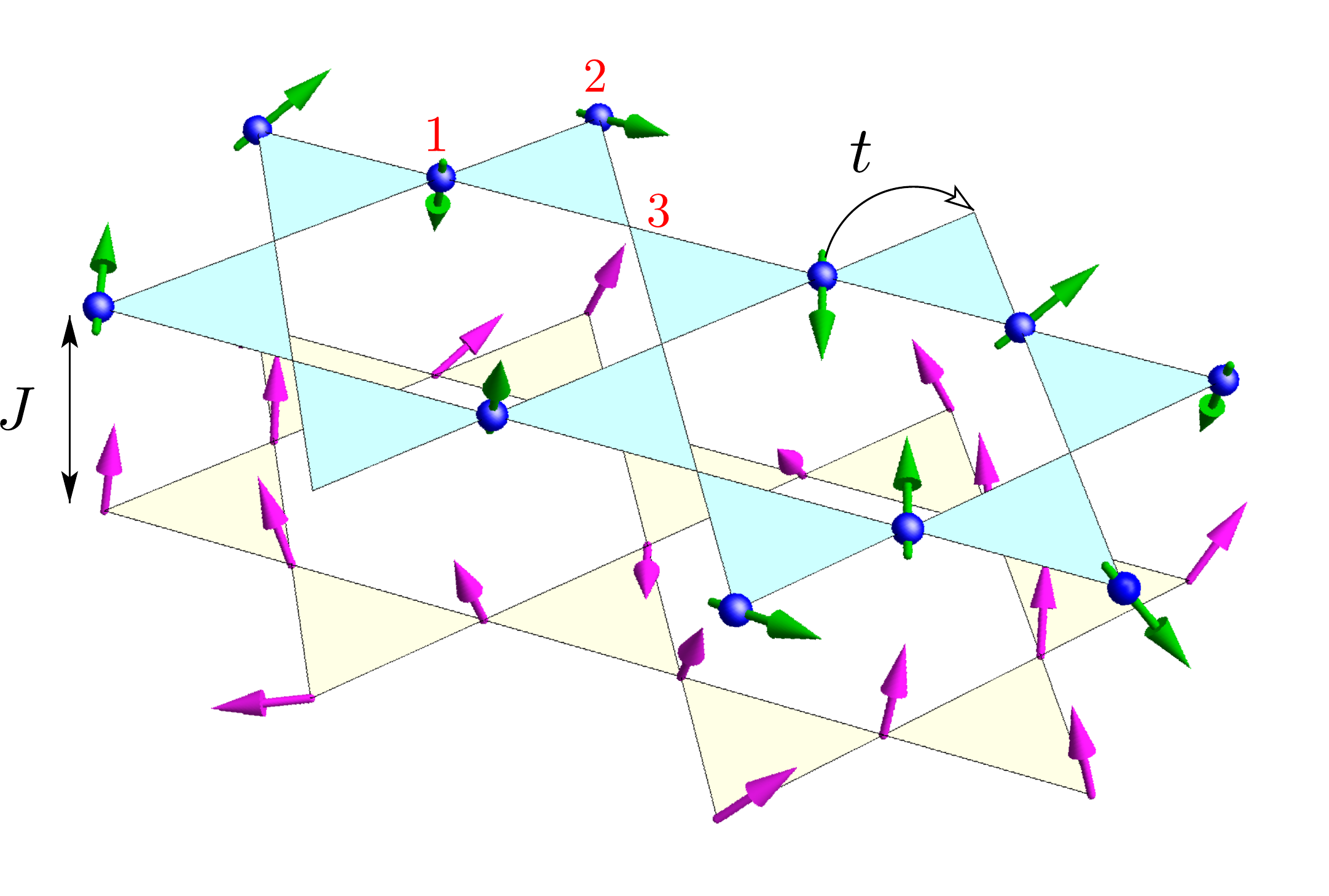}
\caption{\label{fig:KLM} Kondo-lattice model (KLM) on the kagome lattice. The model can be viewed as consisting two subsystems represented here by two fictitious kagome layers. Conduction electrons hop on the top layer according to a nearest-neighbor transfer integral $t$.  The bottom layer consists of localized spins (magenta arrows) at every kagome site. The electron spins (cyan arrows) are coupled to the localized moments through an on-site exchange constant $J$. The numbers 1, 2, 3 denote the three inequivalent sublattices of kagome.}
\end{figure}

In this paper we investigate the Kondo-lattice model (KLM) on the kagome lattice. The KLM is the simplest model that captures the interplay between conduction electrons and localized magnetic moments; see Fig.~\ref{fig:KLM} for an illustration. Because we are interested in cases where the spins of localized electrons develop a net magnetic moment  below a certain temperature ($ \langle {\bf S}_i  \rangle \neq {\bf 0} $ for $T<T_c$), the model can be further simplified by assuming that the localized moments are classical variables, i.e., that there is no Kondo effect. In KLM, conduction electrons interact with the localized spins through an on-site exchange coupling $J$. This exchange corresponds to Hund's coupling for the case of transition metal oxides and Kondo coupling for the case of intermetallics. The magnetic ordering induced by this exchange coupling thus depends on the dispersion of the itinerant electrons at the relevant filling fractions. The kagome lattice has a rich band structure containing Dirac points, quadratic Fermi points, and perfectly nested Fermi surfaces with Van Hove singularities. It is remarkable that a single model contains all of the above features, which have been studied individually in triangular~\cite{martin08,kato10,Akagi10}, honeycomb~\cite{Venderbos11}, checkerboard~\cite{Venderbos12}, cubic~\cite{Hayami14}, and pyrochlore lattices~\cite{chern10,Ishizuka13,Ishizuka13b}. Thus, the kagome KLM is a good candidate to search for robust topologically non-trivial states.

To investigate the complex magnetic orderings of the kagome lattice, our approach includes group-theoretical symmetry analysis at the ordering wavevectors, variational Fourier-space calculations, and large-scale unconstrained numerical simulations. All approaches agree in the small $J$ limit. Our large-scale simulations are based on a recently developed method to simulate classical degrees of freedom interacting with fermions~\cite{barros13}, which enables the unbiased study of magnetic ordering at any $J$. We have uncovered some surprising spin textures that would be very difficult to stabilize in local moment Mott insulating systems. Among them, there are several spin orderings appearing at different filling factors that lead to a spontaneous QAH effect. This variety of magnetic phases opens a path towards the realization of the QAH effect at ambient temperature. 

Our results are also relevant for interacting electron systems without preformed local moments, such as Hubbard-type models~\cite{yu12,wang13,kiesel12,kiesel13}. The kagome tight-binding model is expected to exhibit weak-coupling instabilities when the electron filling reaches the QBCP or the Van Hove singularities. Consequently, an infinitesimal $J$ will immediately produce magnetic ordering. On the other hand, since the Dirac points are stable against weak perturbations, robust magnetic ordering can only be expected at intermediate electron-electron interactions. Moreover, the magnetic or spin-density-wave ordering in the interacting systems has to complete with other instabilities, particularly the superconductivity order~\cite{nandkishore11,yu12,wang13,kiesel12,kiesel13}.  The situation is more complicated for the second Van Hove singularity due to sublattice interference phenomenon~\cite{kiesel12,kiesel13}; we will discuss its consequences in Sec.~\ref{filling2}

The paper is organized as follows. In Section~\ref{modelmethod} we present the KLM on the kagome lattice, discuss the spectral features, and explain our numerical techniques. Next, we present our magnetic ordering results for electron filling fractions $f$ of special interest. In Section~\ref{vanhove} we discuss the case of Van Hove fillings ($f= 3/12$ and $5/12$) in which the Fermi surfaces are nested by three commensurate ordering vectors. In Section~\ref{fermipoints} we discuss the band touching points, including Dirac points ($f=4/12$) and quadratic band crossing ($f=8/12$). Section~\ref{conclusions} summarizes our findings.

\section{Model and methods~\label{modelmethod}}

\subsection{Kondo-lattice model~\label{klm}}

In this work we consider the KLM, which is illustrated in Fig.~\ref{fig:KLM}. Its Hamiltonian is given by
\begin{eqnarray}
\label{eq:H-KLM}
	\mathcal{H} = -t \sum_{\langle ij \rangle} c^{\dagger}_{i\alpha} c^{\;}_{j\alpha} 
	- J \sum_i \mathbf S_i \cdot c^{\dagger}_{i\alpha} \bm\sigma_{\alpha\beta} c^{\;}_{i\beta}\;.
\end{eqnarray}
The first term describes electron hopping between nearest-neighbor (NN) sites; $t > 0$ is the transfer integral, $c^{\dagger}_{i\alpha}$ creates an electron with spin $\alpha$ on site $i$, and $\langle ij \rangle$ denotes a NN pair.  The itinerant electrons interact with the localized spins through an on-site exchange coupling $J$ as described by the second term, where $\bm\sigma_{\alpha\beta}$ is a vector of the Pauli matrices. Note that summation over repeated indices is assumed. We consider the classical limit $|\mathbf S_i| = S \gg 1$ for the localized spins. In this limit, the electron spectrum is independent of the sign of $J$ and the eigenstates corresponding to opposite signs are connected by a global gauge transformation \cite{martin08}.

\begin{figure}
\includegraphics[width=0.9\columnwidth]{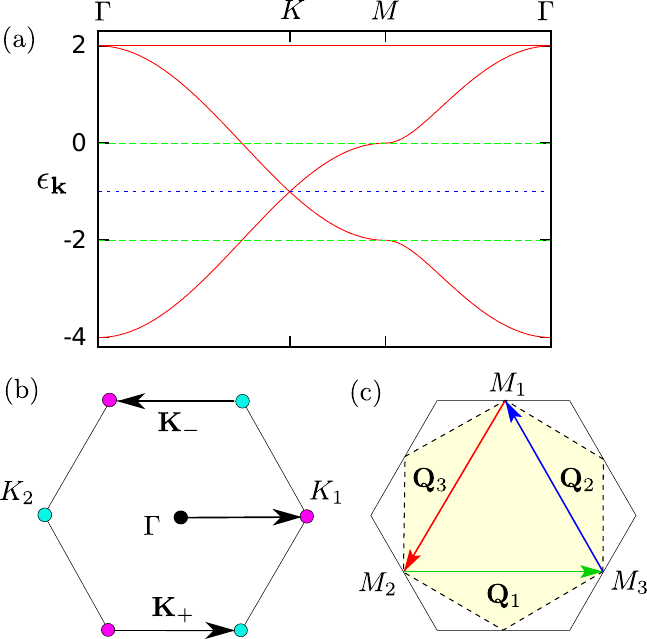}
\caption{\label{fig:dispersion} (a) Band structure of the tight-binding model on the kagome lattice. Here and in the following figures, the band energy $\epsilon_{\mathbf k}$ is measured in units of NN hopping $t$. (b) Brillouin zone (BZ) of the kagome lattice. At filling fraction $f=1/3$ the Fermi surface shrinks to two inequivalent Dirac points $K_{1,2}$ at the corners of the BZ. These two Dirac points are connected by wavevectors $\mathbf K_{\pm} = (\pm 4\pi/3, 0)$. (c) The dispersion has three saddle points $M_{1,2,3}$ at the edge of the BZ, giving rise to a logarithmically divergent DOS at filling fractions $f = 1/4$ and $5/12$. The corresponding Fermi surface is a regular hexagon inscribed within the BZ. Pairs of parallel edges of this Fermi surface are perfectly nested by wavevectors $\mathbf Q_1 = (2\pi, 0)$ and $\mathbf Q_{2,3} = (-\pi, \pm \sqrt{3}\pi)$.}
\end{figure}

The energy dispersion for the kagome lattice tight-binding model, i.e. the first term in Eq.~(\ref{eq:H-KLM}), can be obtained analytically. We first label the kagome sites as $i =({\mathbf r}_i )= (\mathbf r, m)$, where ${\mathbf r}_i$ denotes the position of site $i$, which is decomposed into the triangular Bravais lattice point ${\mathbf r}$ and the sublattice index $m = 1,2,3$; see Fig.~\ref{fig:KLM}.  After Fourier transformation $c_{i,\alpha} = (1/\sqrt{N})\sum_{\mathbf k} c_{m,\alpha}(\mathbf k) \exp(i\mathbf k\cdot {\mathbf r}_i)$, we have $\mathcal{H}_t = \sum_{m,n=1}^3 h_{mn}(\mathbf k)\,c^\dagger_{m\alpha}(\mathbf k) \,c^{\;}_{n\alpha}(\mathbf k)$. The hopping matrix is
\begin{eqnarray}
	\label{eq:hopping}
        \hat h(\mathbf k) = -2 t \, \left(\begin{array}{ccc}  0 & \cos\frac{\mathbf k\cdot\mathbf a_2}{2} & \cos\frac{\mathbf k \cdot \mathbf a_1}{2} \\
	\cos\frac{\mathbf k \cdot\mathbf a_2}{ 2} & 0 & \cos\frac{\mathbf k \cdot\mathbf a_3}{2} \\
	\cos\frac{\mathbf k \cdot \mathbf a_1}{2} & \cos\frac{\mathbf k \cdot\mathbf a_3}{2} & 0 \end{array} \right)\;,
\end{eqnarray}
where $\mathbf a_1 = (1,0)$, $\mathbf a_2 = (1/2,\sqrt{3}/2)$ are primitive lattice vectors of the triangular Bravais lattice, and $\mathbf a_3 = \mathbf a_2 - \mathbf a_1$. After diagonalizing this matrix, we obtain two dispersive bands 
\[
\epsilon_{1,2} = -t \mp t\,\sqrt{3+2 \, \textstyle{\sum_{\nu=1}^3}\cos\mathbf k\cdot\mathbf a_{\nu}}\;,
\] 
and a flat band $\epsilon_{3} = 2\, t$ at the top of the dispersion.  

As mentioned in the introduction, the kagome lattice electronic band structure [Fig.~\ref{fig:dispersion}(a)] exhibits several special points, which we now present and discuss in more detail. First, there are two inequivalent Dirac cones at the corners $K_{1,2}$ of the Brillouin zone (BZ) schematically represented in Fig.~\ref{fig:dispersion}(b). As these Dirac points are the isolated crossing points of the two dispersive bands,  the Fermi surface shrinks to the two $K$--points at filling fraction $f = 1/3$. In accordance with the standard theory of electronic Berry phases, these Dirac Fermi points are characterized by nonzero Berry flux $\pm \pi$~\cite{Haldane04}. In the case of the kagome lattice, it has been shown that gapping out a pair of Dirac points can lead to topological phases such as spontaneous QAH and quantum spin Hall insulators~\cite{ohgushi00,guo09}, in agreement with the general theory of gapped Dirac fermions in two dimensions~\cite{Haldane88,Kane05}. However, the vanishing density of states (DOS) renders the Dirac Fermi points robust against weakly repulsive electron-electron interactions. In general, a finite interaction stength is required to open a gap at $f = 1/3$ and stabilize a topologically nontrivial phase~\cite{liu10,wen10}.

Another topologically nontrivial band-touching occurs at the $\Gamma$-point between the upper dispersive and flat bands. At filling fraction $f=2/3$ the Fermi surface consists of a single point at $\mathbf k = \mathbf 0$ with a quadratic dispersion in its vicinity: $\epsilon \sim k^2$. This so-called quadratic band-crossing point (QBCP) is characterized by a $\pm 2\pi$ Berry phase. In momentum space, the topological QBCP resembles a vortex with a $\pm 2$ winding number~\cite{sun09,volovik-book}. Contrary to Dirac points, quadratic Fermi points are unstable against arbitrarily weak short-range interactions due to their finite DOS~\cite{sun09,liu10}. The perturbed quadratic Fermi point either splits into two fundamental Dirac points or it is completely gapped out, giving rise to a topological insulator~\cite{sun09}. However, because the whole flat band $\epsilon_3$ is degenerate with the $\Gamma$ Fermi point, finding the magnetic structure stabilized by $\mathcal{H} $ in Eq.~(\ref{eq:H-KLM}) for $f = 2/3$ is still a rather complicated problem.

A third distinct spectral feature is related to the two dispersive bands, $\epsilon_1$ and $\epsilon_2$. The DOS of $\epsilon_1$ and $\epsilon_2$ contains a Van Hove singularity at fillings $f = 1/4$ and $5/12$, respectively. A logarithmically divergent DOS at these filling fractions results from the three saddle points $M$ at the edges of the BZ [Fig.~\ref{fig:dispersion}(c)]. For just NN hopping, the Fermi surface at these filling fractions is a regular hexagon inscribed within the BZ, as shown in Fig.~\ref{fig:dispersion}(c). Remarkably, pairs of parallel edges of this Fermi surface  are perfectly nested by wavevectors $\mathbf Q_1 = (2\pi, 0)$ and $\mathbf Q_{2,3} = (-\pi, \pm \sqrt{3}\pi)$. The perfect Fermi surface nesting combined with a divergent DOS leads to a magnetic susceptibility of the conduction electrons that diverges as $\log^2{|{\bf k}-{\bf Q}_\eta}|$ ($\eta=1,2,3$). Consequently, the system is unstable against developing a triple-$\mathbf Q$ magnetic order even for small $J/t$.

Our goal is to find the magnetic ground state of the KLM for each of these special filling fractions and to establish the nature of the corresponding electronic states. In the process of uncovering the magnetic ordering at the special fillings mentioned above, we will also discuss specific fillings different from these three. A quadrupling of the unit cell as a consequence of finite ${\bf Q}_\eta$ ordering allows for commensurate fillings other than the Van Hove fillings $f=1/4$ and $f=5/12$, which are $f=1/12$ and $f=7/12$. We have investigated the magnetic ordering and the corresponding electronic state at these fillings as well. 

\subsection{Numerical methods}

In a na\"ive approach to Monte Carlo sampling of the classical spins, every trial update of the spin configuration requires diagonalization of a new single-particle electron matrix to determine the change in energy. Direct diagonalization costs order $N^3$ numerical operations, so this simulation approach is limited to systems with $N \approx 16^2$ spins (e.g.  Ref.~\onlinecite{kato10}). We require much larger scale systems to study weak-coupling instabilities. Small systems obscure susceptibility divergences because of their larger momentum space discretization cut-off. For example, Fermi surface nesting yields a divergent susceptibility that scales as $(\log N)^2$, and stabilizing the desired triple-$\mathbf Q$ magnetic orders may require $N \approx 100^2$ spins~\cite{barros13}. Evidently, direct diagonalization at every Monte Carlo step is impractical. 

To simulate the KLM at large scales, we employ two complementary numerical methods. First, we perform unconstrained, finite-temperature simulations with a recently developed Langevin method, which we review below. Second, we use a Fourier-space variational method to verify and analyze the candidate orderings at zero temperature. In our variational approach, we assume a known magnetic unit cell size and search for the lowest energy ordering among spin configurations within the extended unit cell~\cite{chern10}. For example, the most general triple-$\mathbf Q$ orderings have a quadrupled magnetic unit cell of 12 spins.  The electronic part can then be solved by applying a Fourier transformation. The total energy obtained by summing over states within the reduced BZ is a function of the magnetic configuration in the extended unit cell. We use standard simulated annealing to find the minimum-energy configurations with constrained periodicity~\cite{chern10}. More details can be found in Appendix~\ref{appendix1}.

Our Langevin method enables high accuracy, unconstrained simulation of very large systems, and is orders of magnitude faster than previous Monte Carlo methods~\cite{barros13}. We approximate Langevin forces using an extension of the kernel polynomial method (KPM)~\cite{silver94,wang94,weisse06}. KPM provides fast estimates of the density of electron states (DOS), $\rho(\epsilon) = \sum_m \delta(\epsilon - \epsilon_m[\mathbf S_i])$, where $\epsilon_m$ are the electron energy levels, i.e. the eigenvalues of the single-particle electron matrix $H$. KPM approximates the DOS using a truncated series in Chebyshev polynomials $T_m(\epsilon)$,
\[
\rho(\epsilon) \approx \frac{1}{ \pi \sqrt{1-\epsilon^2} } \sum_{m=0}^{M-1} (2 - \delta_{0,m}) g_m \mu_m T_m(\epsilon)\;.
\]
The numerical factors $g_m$ associated with the Jackson kernel~\cite{jackson} optimally damp Gibbs oscillations~\cite{weisse06}. The Chebyshev moments are given by $\mu_m = {\rm Tr}\, T_m(H)$. For efficiency, we use the KPM stochastic approximation $\mu_m \approx r^\dagger \cdot T_m(H) r$, where $r$ is a random vector with elements that satisfy the ensemble average $\langle r_i^\ast r_j\rangle = \delta_{i,j}$. We use the Chebyshev recursion relation in the form $\alpha_m = 2 H \alpha_{m-1} - \alpha_{m-2}$ to iteratively build the vectors $\alpha_m = T_m(H) r$. Thus, the moment estimates $\mu_m \approx r^\dagger \cdot \alpha_m$ may be evaluated at linear cost in the number of spins $N$ by using only sparse matrix-vector multiplications. KPM accuracy is controlled by the  series truncation order $M$ and the number of random vectors $R$ over which we average.

The KPM DOS allows estimation of the (free) energy $\mathcal{F}$  of a spin configuration $\{\mathbf S_i\}$. After integrating out the electrons at fixed chemical potential $\mu$ and inverse temperature $\beta = 1/k_B T$ we obtain
\[
\mathcal{F}[\mathbf S_i] = -\beta^{-1} \log {\rm Tr}_{c} \,e^{-\beta (\mathcal{H} - \mu\sum_i c^{\dagger}_i c^{\,}_i)} 
	=  \int \rho(\epsilon) f(\epsilon) d\epsilon \;,
\]
where $f(\epsilon) = -\beta^{-1} \log \{1 + \exp[-\beta (\epsilon-\mu)]\}$ is an anti-derivative of the Fermi function. Chebyshev-Gauss quadrature allows fast and accurate numerical integration over the KPM DOS estimate~\cite{weisse06}.

To efficiently sample spin configurations $\{\mathbf S_i\}$ we extend KPM to also estimate forces $-\partial \mathcal F / \partial \mathbf S_i$. One path is the numerically exact technique of automatic differentiation with reverse accumulation~\cite{griewank}. Indeed, {\it all} forces $-\partial \mathcal F / \partial \mathbf S_i$ may be simultaneously estimated by a ``reverse'' recursion relation with a cost linear in $N$ that is equivalent to the cost of estimating $\mathcal F$~\cite{barros13}.

We apply our KPM based force estimates to sample spins according to overdamped Langevin dynamics,
\begin{eqnarray}
	\label{eq:langevin}
	\mathbf S_i(\tau+\Delta \tau) - \mathbf S_i(\tau) = -\Delta \tau \frac{\partial\mathcal{F}}{\partial \mathbf S_i} 
	+ \sqrt{2\,\beta^{-1}\Delta \tau}\,\,\bm\eta_i(\tau) \;,
\end{eqnarray}
where $\bm\eta_i(\tau)$ are uncorrelated Gaussian random variables with unit variance, $\tau$ is the Langevin time, and we use an implicit Lagrange multiplier to constrain $\partial |\mathbf S| / \partial \tau = 0$. Accuracy is again controlled by two tunable parameters: the series truncation order $M$ and the fraction $z = \Delta \tau / R$ of Langevin integration time per random vector. The cost to integrate one unit of Langevin time scales as $N M / z$. KPM inaccuracies may be viewed as introducing an effective temperature. Series truncation effectively smooths the Fermi function on the scale $\Delta T_1 \sim 1/M$, whereas stochastic errors in the force effectively increase the thermal noise an amount $\Delta T_2 \sim (J/t)^2 z$. To search for ground states, we randomize the initial spin configuration and then integrate the Langevin dynamics with $\beta^{-1} = 0$.  We perform most simulations with $M=500$ and $z = 0.02$. The effective accuracy improves with decreasing exchange coupling. At $J/t=0.1$ we can often distinguish between spin textures with energies that differ at the fifth significant digit.

Because the Langevin dynamics Eq.~(\ref{eq:langevin}) can be viewed as the overdamped limit of the Landau-Lifshitz-Gilbert equation, our simulations also capture the physical emergence of mesoscale topological defects and their dynamics. Indeed, we find different domain structures in the same magnetic ordering for different filling fractions at finite temperatures. This subtle difference can be attributed to the different types of effective long-range spin-spin interactions that are mediated by the conduction electrons.

\section{Van Hove Singularities~\label{vanhove}}

We start by considering the magnetic orderings for electron filling fractions right at the Van Hove singularities of the DOS, i.e. the filling fractions $f=3/12=1/4$ and $f=5/12$. As mentioned above, the Fermi surface of the ideal NN kagome lattice tight-binding model is a regular hexagon inscribed within the hexagonal BZ. The corners of the Fermi surface hexagon are the three inequivalent $M$-points of the BZ. These special points are the dominant source of the divergent susceptibility as they are the saddle points of the electron dispersion. The $M$-points are nested by three commensurate wavevectors $\mathbf Q_1 = (2\pi, 0)$, and $\mathbf Q_{2,3} = (-\pi, \pm \sqrt{3}\pi)$, which generically give rise to a quadrupled unit cell since these nesting wavevectors are half of the reciprocal lattice vectors. Before we present the results of our numerical simulations, we will provide a symmetry-based perspective of general triple-$\mathbf Q$ magnetic orderings, with the purpose of gaining insight into the magnetic and electronic properties of such orderings. 

In addition, we will demonstrate how the magnetic ground states found in the numerical simulations may be understood analytically on the basis of symmetry constraints~\cite{Venderbos141}. We then move on to discuss in detail the magnetic order parameters describing the spin textures obtained from our numerical minimizations for $f = 3/12$ and $f=5/12$, respectively.

\subsection{Symmetry properties of triple-Q orderings~\label{symmetry}}

Triple-$\mathbf Q$ magnetic ordering on the kagome lattice can be discussed from the perspective of lattice symmetries. This will shed light on the electronic properties of the conduction electrons in the presence of such ordering. Furthermore, we will illustrate how triple-$\mathbf Q$ magnetic ground states can be derived systematically by imposing symmetry requirements on the most general form of the spin order parameter~\cite{Venderbos142}. The numerical simulations then confirm that these are the correct requirements to impose. 

As a consequence of the commensurability of the $M$-point wavevectors we have $2\mathbf Q_\eta \equiv \mathbf 0$ (modulo reciprocal lattice vectors) and hence $\cos(\mathbf Q_\eta\cdot \mathbf r) = e^{i \mathbf Q_\eta\cdot \mathbf r} = \pm 1$.  The most general spin state modulated by these $\mathbf Q_\eta$ vectors  can be written as 
\begin{equation}
	\mathbf S_i = \mathbf S_m(\mathbf r) = \textstyle{\sum_{\eta = 1}^3} \bm\Delta^m_{\eta}\, \cos(\mathbf Q_\eta\cdot\mathbf r) \;,
\end{equation}
where every site on the kagome lattice is labeled by $i = (m, \mathbf r)$, $m$ denoting the sublattice and $\mathbf r = l \mathbf a_1 + n \mathbf a_2$  denoting the Bravais lattice unit cell. The set $\{\bm\Delta^m_{\eta}\}$ describes nine vector order parameters, one for each sublattice and $\mathbf Q$-vector. 

In the context of the KLM with classical spin states, we should constrain the vector order parameters $\{\bm\Delta^m_{\eta}\}$ to configurations that satisfy: $|\mathbf S_i| = S$, i.e., equal spin length on every site. For the moment we will ignore this constraint and work with the most general set of order parameter components, which is captured by $\{\bm\Delta^m_{\eta}\}$. The constraint will be reinstated in a natural way later. We may write the vector order parameters a product of a scalar and a vector with unit length, i.e. $\zeta^m_\eta\, \hat{\mathbf n}_\eta^m$.  Focusing first exclusively on the scalar part, we can organize them in terms of distinct representations of the lattice symmetry group. For simplicity, we group the scalar order parameters into three vectors $\{\vec\zeta_{\eta}\}$ which together form a 9-dimensional representation of the lattice symmetry group. Its irreducible representations describe possible site orderings or density-wave states on the kagome lattice.

\begin{figure}
\includegraphics[width=0.99\columnwidth]{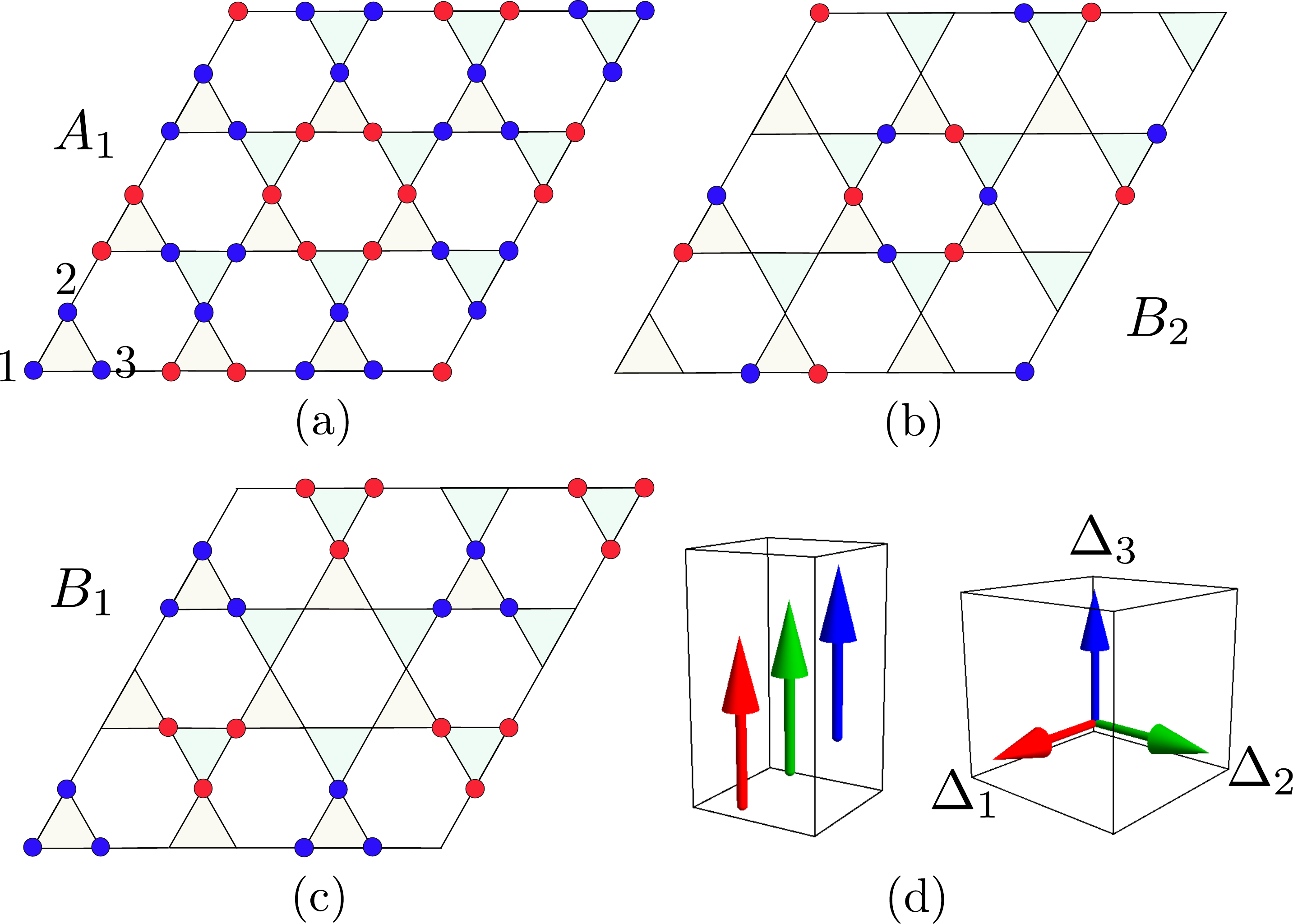}
\caption{\label{fig:irrep} (a)--(c) Schematic diagram showing the triple-$\mathbf Q$ scalar order (or collinear spin order) corresponding to the three irreducible representations $A_1$, $B_2$, and $B_1$. The blue (red) circles denote on-site scalar variable $\zeta_i = +1$ ($-1$), while empty site corresponds to $\zeta_i = 0$. Panel (d) shows the collinear and orthogonal configurations of the triple-$\mathbf Q$ vector order parameters.
}
\end{figure}

The point group of the kagome lattice is $C_{6v}$. However, each ordering vector $\mathbf {Q}_\eta$ is left invariant by the subgroup $G_{\mathbf Q_{\eta}}=C_{2v}$, called the small group of the ordering vectors.  
The latter admits only one-dimensional ($1D$) irreducible representations. Taking one of the ${\bf Q}_\eta$, for instance ${\bf Q}_3$, we can evaluate the constraints of the symmetry elements in $C_{2v}$ on $\vec{\zeta}_3$. Doing so yields three solutions for  $\vec{\zeta}_3$, which is consistent with the three sublattices of the kagome lattice. The solutions correspond to the irreducible representations $A_1$, $B_1$ and $B_2$ of $C_{2v}$. Having found the solutions for one ${\bf Q}_\eta$, we obtain the solutions for the others by (rotational) symmetry. In total we have nine scalars $\zeta^m_\eta$ and for each of the three solutions for a single ${\bf Q}_\eta$ we can form a symmetric combination and the two $d$-wave combinations, yielding nine states in total. The symmetric combinations will transform as a $1D$ irreducible representation of $C_{6v}$, which are $A_1$, $B_1$ and $B_2$. These three states are the relevant building blocks for the spin ordered states. 

The explicit form of the $\zeta^m_\eta$ corresponding to the three $1D$ representations can be determined by explicitly evaluating the symmetry operations of the $C_{2v}$ group. For instance, the two-fold rotation requires  $C_{2}\,\vec\zeta^{A_1} = + \vec\zeta^{A_1}$, and $C_2\,\vec\zeta^{B} = -\vec\zeta^{B}$. The two $B$ representations may be distinguished by their transformation properties under reflection. Explicit calculations show that
\begin{eqnarray}
	\vec\zeta^{A_1}_3 = \left(\begin{array}{c} 1 \\ 0 \\ 0 \end{array}\right), \quad
	\vec\zeta^{B_1,B_2}_3 = \frac{1}{\sqrt{2}}\left(\begin{array}{c} 0 \\ 1 \\ \pm 1 \end{array}\right), 
\end{eqnarray}
where the $+$ or $-$ signs correspond to the $B_1$ or $B_2$ modes, respectively.  Once $\vec\zeta_3$ is determined, the other two vectors can be obtained by applying the three-fold rotations. The pattern of the site-ordered state corresponding to the above three irreducible representations is shown in Fig.~\ref{fig:irrep}(a)--(c). These also correspond to the spin configurations in collinear SDW states.

We take these three symmetric combinations and focus on the different ways of embedding the obtained $\{\vec\zeta_\eta\}\;  = \{\zeta^m_\eta\}$  in a vector order parameter, i.e. the different ways of combining them with $ \mathbf {n}_\eta^m$. A fully collinear spin state would be given by $\mathbf {n}_\eta^m = \mathbf {n}$ for all sublattices $m$, and nesting wavevectors $\mathbf Q_\eta$. This fully collinear arrangement of spins does not change the symmetry of the electronic state, which is still $A_1$, $B_1$ and $B_2$, respectively. Due to the fact that the collinear $A_1$ state uses only a single $\mathbf {Q}_\eta$ per kagome sublattice, it fulfills the requirement of uniform spin length, whereas the (collinear) $B_{1,2}$ states do not. We note that the $A_1$ state would be the kagome lattice version of the uniaxial spin density wave state reported in \cite{nandkishore12,chern12,chern12b}. These collinear spin states all manifestly break translational invariance as a consequence of ordering at finite $\mathbf {Q}_\eta$ momentum vectors. 

One can restore an effective translational invariance by choosing $\mathbf {n}_\eta^m = \mathbf n_\eta$, i.e. single unit vector for each $\mathbf Q_\eta$, while at the same time demanding $\mathbf {n}_\eta \perp \mathbf {n}_\nu $ for $\eta\neq \nu$ [see Fig.~\ref{fig:irrep}(d)]. Translational invariance is effectively preserved for these spin configurations, as the translations which appear to be broken by finite $\mathbf Q_\eta$ ordering can be combined with global $O(3)$ rotations of the spins to leave the state invariant (see also Ref.~\cite{messio11}). For each of the three scalar states there is such a corresponding translationally invariant spin state. Because these spin states are translationally invariant, they clearly satisfy the constraint of uniform spin length. A key result of our numerical work, which we discuss below, is that these three triple-$\mathbf Q$ spin states are indeed the ground states (modulo a global spin rotation) at the different commensurate filling fractions. 

We summarize the symmetry constraints giving the three noncoplanar spin configurations as follows: the electronic state i) transforms as a $1D$ irreducible representation of $C_{6v}$, and ii) is translationally invariant up to global spin rotation.  

Knowing the symmetry properties of the triple-$\mathbf Q$ spin configurations puts us in a position to immediately deduce the symmetries of the electronic state. First, the effective translational invariance mandates a full double degeneracy of the spectrum. Translating the spin configuration by one lattice vector and performing a global spin rotation by $\pi$ about the appropriate axis does not change the Hamiltonian, but it does transform the wave function into an orthogonal one. Hence, the electronic spectrum is manifestly doubly degenerate for all three states. 

By associating a distinct orthogonal spin component $\mathbf {n}_\eta$ to each of the three $\mathbf {Q}_\eta$ momenta [Fig.~\ref{fig:irrep}(d)], the resulting triple-$\mathbf Q$ states have noncoplanar spin configurations. Up to a global $O(3)$  spin rotation, these magnetic states are invariant under the reflection operations of the $C_{6v}$ group. Specifically, one can show that such a rotation is improper, i.e., it gives a minus sign when translated into an $SU(2)$ rotation on the electron spin. This changes the symmetry of the electronic state to  $A_2$, $B_2$ and $B_1$, respectively, after multiplication by $A_2$. Therefore, the electronic states corresponding to the three triple-$\mathbf Q$ spin ordered states transform as the representations $A_2$, $B_2$ and $B_1$ of the hexagonal symmetry group $C_{6v}$.

The noncoplanarity of the spins admits a discrete scalar order parameter characterizing the chirality of the structure. Explicitly, a scalar chiral order parameter can be defined for each sublattice
\begin{eqnarray}
	\kappa = \bm\Delta_{1} \cdot\bm\Delta_{2} \times \bm\Delta_{3} \;,
\end{eqnarray}
where $\Delta_\eta$ refers to ordering vector ${\bf Q}_\eta$. In general, the $\mathbb Z_2$ chiral order parameters of different sublattices are related to each other. More importantly, although the continuous spin $SU(2)$ symmetry cannot be spontaneously broken in 2D,  true long-range ordering of the discrete chirality is possible. Consequently, one expects a finite temperature phase transition that breaks the $\mathbb Z_2$ symmetry. Indeed, a similar transition is observed in the tetrahedral ordering transition in the triangular KLM~\cite{kato10,barros13}. 

In the rest of this section we will discuss the ground states observed in numerical simulations, which we find to be consistent with the predictions based upon symmetry analysis.
We stress that our large-scale KPM-Langevin numerical simulations make no {\it a priori} assumptions concerning the magnetic ordering. Our KPM-Langevin results are also confirmed by the variational Monte Carlo method in Fourier space; see Appendix~\ref{appendix1} for more details.

\begin{figure}
\includegraphics[width=0.88\columnwidth]{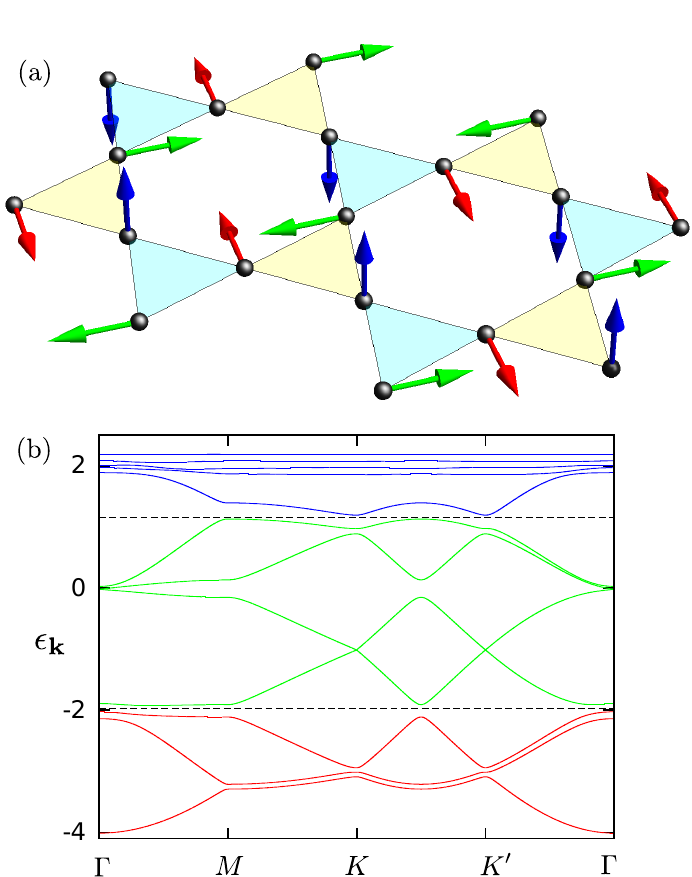}
\caption{\label{fig:vH1} (a) The triple-$\mathbf Q$ noncoplanar (orthogonal) spin order on kagome lattice that transforms according to the $A_2$ irreducible representation of the symmetry group. Spins at the three inequivalent sublattices are indicated by different colors. This $A_2$ symmetry   state is the ground state at filling fraction $f = 1/4$ for small coupling and $f = 7/12$ for intermediate coupling ($J \gtrsim 0.2$).   The corresponding band structure is shown in panel (b). The exchange coupling $J = 0.22\, t$ is used in the calculation. The dashed lines indicate the Ferme levels at filling fractions $f = 1/4$ and $7/12$.
}
\end{figure}

\subsection{Van Hove filling $f = 1/4$~\label{filling1}}
\label{sec:f1_4}

\begin{figure}[t]
\includegraphics[width=0.99\columnwidth]{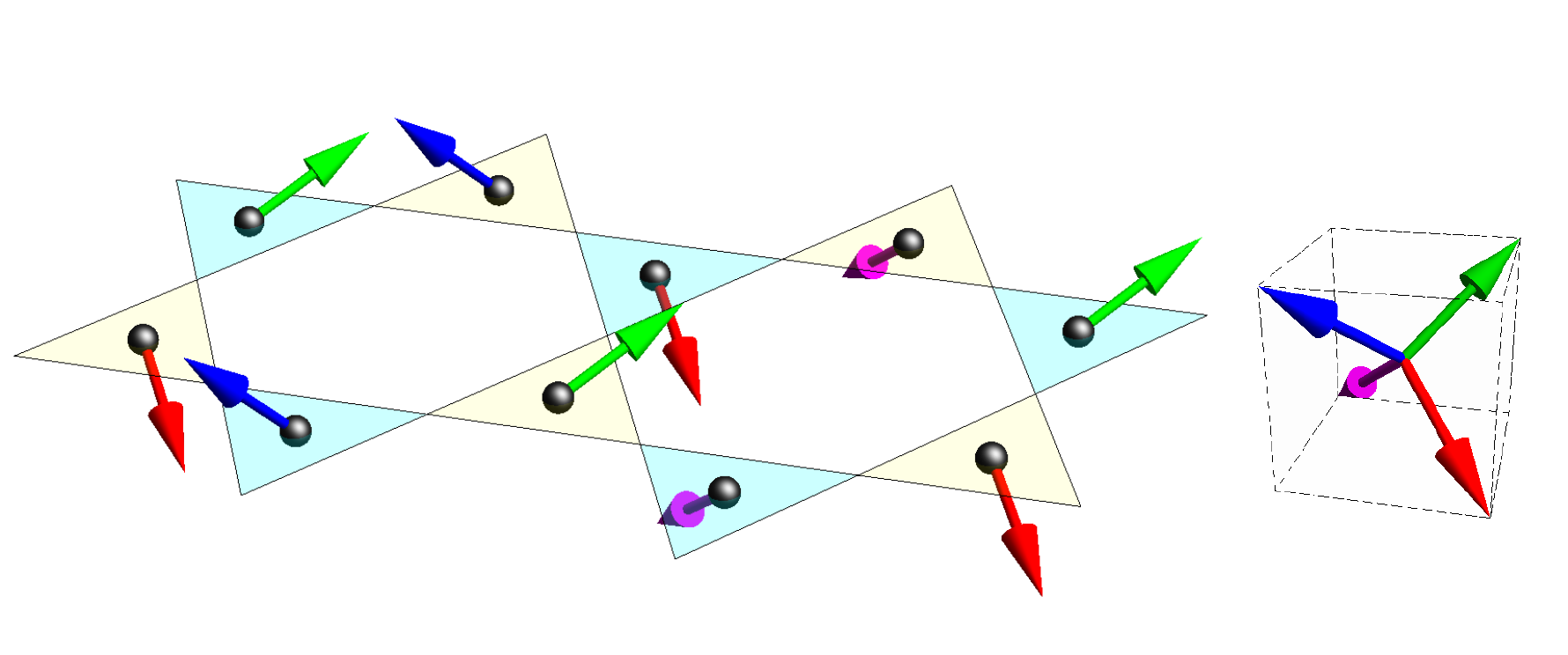}
\caption{\label{fig:chiral} The chiral structure of spin orders at Van Hove fillings in the kagome lattice. There is a vector associated with each triangle. The four different orientations of the vectors (indicated by four different colors) point to the corners of a regular tetrahedron. The vectors associated with each triangle correspond to the total spins $\mathbf M_{\triangle} = \mathbf S_1 + \mathbf S_2 + \mathbf S_3$ in the $A_2$ symmetry order at $f=1/4$ filling [Fig.~\ref{fig:vH1}(a)], and to the vector chirality $\bm\chi_{\triangle} =\mathbf S_1\times\mathbf S_2 + \mathbf S_2 \times \mathbf S_3 + \mathbf S_3 \times \mathbf S_1$ in the $B_2$ symmetry order at the second Van Hove filling $f = 5/12$ [Fig.~\ref{fig:vH2}(a)]. }
\end{figure}

\begin{figure*}[t]
\includegraphics[width=1.98\columnwidth]{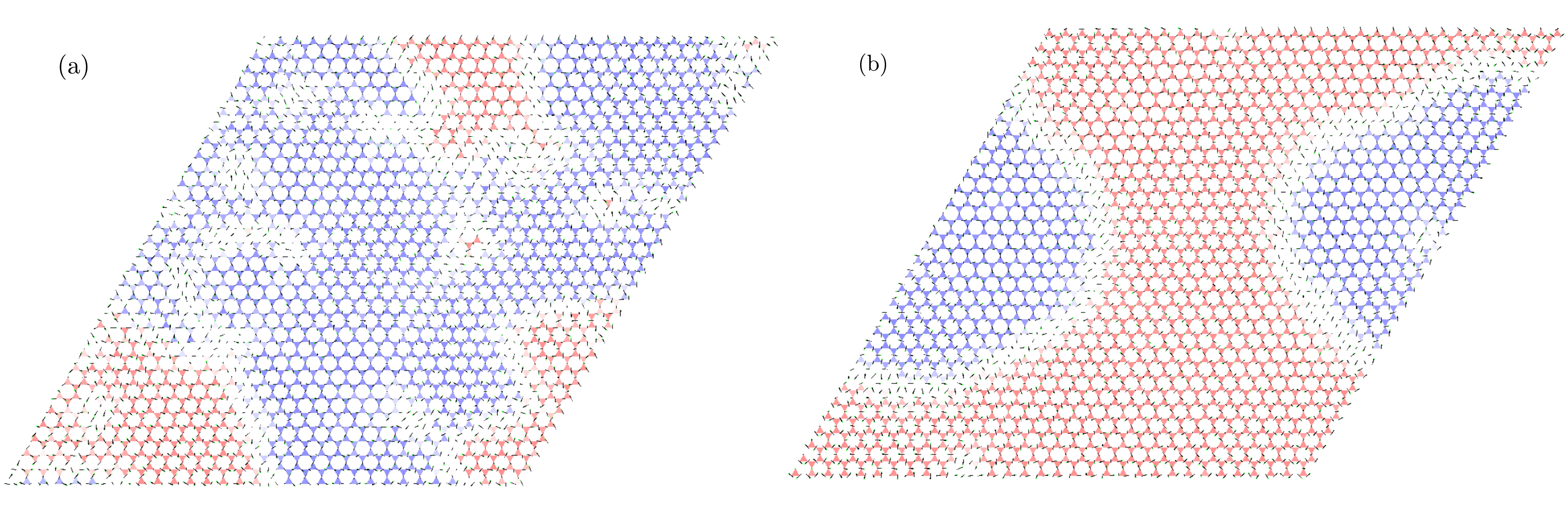}
\caption{\label{fig:snapshot1} Snapshots of the scalar chirality field obtained from KPM-Langevin simulations of $N = 3 \times 32^2$ spins at filling fractions (a) $f = 1/4$ and (b) $f = 7/12$. The magnetic ground state at these two filling fractions is the $A_2$ symmetry state shown in Fig.~\ref{fig:vH1}(a). At the intermediate times shown, nonequilibrium features such as chiral domain walls and $\mathbb Z_2$ vortices are apparent, although they will eventually annihilate under the (nearly) zero temperature KPM-Langevin dynamics. We use a moderate coupling constant of $J = 0.5\,t$ for these simulations. 
}
\end{figure*}

The first filling fraction we discuss is the Van Hove filling $f=3/12=1/4$. The spin structure obtained from KPM-Langevin simulations with $0 \leq J \leq 2 t$ is the $A_1$-symmetry state shown in Fig.~\ref{fig:vH1}(a). Our variational Fourier-space minimization also confirms this spin order is the minimum-energy state at $f=1/4$.  The nonzero order parameter components at the three nesting wavevectors are:
\begin{eqnarray}
	\bm\Delta^3_{1} = S\, \hat{\mathbf n}_1 \;, \quad
	\bm\Delta^2_{2} = S\, \hat{\mathbf n}_2 \;, \quad
	\bm\Delta^1_{3} = S\,\hat{\mathbf n}_3 \;. 
\end{eqnarray}
Here \{$\hat{\mathbf n}_1$, $\hat{\mathbf n}_2$, $\hat{\mathbf n}_3$\} are three arbitrary orthogonal unit vectors. In this magnetic order, spins of the same sublattice are collinear with respect to each other, $\mathbf S_1(\mathbf r) = \pm S\, \hat{\mathbf n}_3$, $\mathbf S_2(\mathbf r) = \pm S\,\hat{\mathbf n}_2$, and $\mathbf S_3(\mathbf r) = \pm S\, \hat{\mathbf n}_1$, while spins at different sublattices are orthogonal to each other. The electronic bands for this magnetic ordering are shown in Fig.~\ref{fig:vH1}(b). The band structure shows twelve bands, consistent with the requirement of (at least) double degeneracy due to a combined translational and rotational symmetry. We observe the opening of a spectral gap at $f = 1/4$ filling for this noncoplanar order. We may view the three sublattices as utilizing three different nesting wavevectors $\mathbf Q_\eta$ to gap out the full Fermi surface.

A noncoplanar magnetic structure implies nonzero chiral order $\kappa  = \bm\Delta_{1} \cdot\bm\Delta_{2} \times \bm\Delta_{3} \sim \hat{\mathbf n}_1 \cdot \hat{\mathbf n}_2   \times \hat{\mathbf n}_3 $. The chiral structure of the noncoplanar order shown in Fig.~\ref{fig:vH1}(a) becomes apparent by considering the effective magnetic moment of each individual triangle. By defining the total spin for individual triangles as $\mathbf M_{\triangle} = \mathbf S_1 + \mathbf S_2 + \mathbf S_3$, we observe that this vector sum at the four inequivalent triangles points to the corners of a regular tetrahedron (Fig.~\ref{fig:chiral}). More specifically, the scalar chirality $\kappa_{\triangle} = \mathbf S_1\cdot\mathbf S_2 \times \mathbf S_3$  is nonzero in each triangle, deriving from a nontrivial solid angle subtended by the three spins. This implies that the electrons hopping on a triangle acquire a nonzero Berry phase in the presence of such noncoplanar order. We therefore examine the electronic state in more detail.

The $A_2$ symmetry spin order itself is found to preserve all rotations and reflections of the hexagonal group. The electronic state however, owing to the finite chirality ($\kappa \neq 0$), transforms as $A_2$, breaking the reflection symmetries. In addition, time-reversal symmetry inverts all spins, an effect which cannot be compensated by a rotation in case of a chiral spin configuration. We coclude that the electronic state admits a topological characterization in terms of the Chern number~\cite{tknn}. Explicit evaluation shows that the Chern number is nonzero for the fully gapped electronic state at $f=1/4$. Consequently, the electronic state corresponds to a QAH state (a Chern insulator). The noncoplanar nature of this spin configuration implies that it breaks the discrete chiral part of the full $O(3)$ symmetry ($O(3)  = \mathbb Z_2 \times SO(3)$). We note again that while the continuous $SU(2) \cong SO(3) $ spin symmetry cannot be broken at finite temperatures in 2D, a discrete $\mathbb{Z}_2$ order parameter {\it can} develop long-range order at finite temperatures. One therefore expects a finite temperature phase transition associated with the chiral symmetry breaking.

The magnetic order at $f=1/4$ filling can also be understood by a simple analysis of electron eigenstates at the three saddle points. We express the electron operators in terms of quasi-particle ones  $c_{i, \alpha} = (1/\sqrt{N}) \sum_{\mathbf k} w^m_{\mathbf k}\, f_{\mathbf k, \alpha} \exp({i\mathbf k\cdot\mathbf r})$, where again the lattice site is labelled by $i = (m, \mathbf r)$, $m = 1,2,3$ denotes the different sublattices and $\mathbf r$ indicates the Bravais lattice point. $f^\dagger_{\mathbf k,\alpha}$ is the quasi-particle creation operator with momentum $\mathbf k$ and $\vec w_{\mathbf k} = \{w^m_{\mathbf k} \}$ is the eigenvector of the hopping matrix, i.e. $ h_{mn}(\mathbf k) \, w^n_{\mathbf k} = \epsilon_{\mathbf k}   w^m_{\mathbf k}$; it contains the sublattice weights of the quasi-particle. Denoting the amplitude of the order parameters as $\Delta$, the exchange coupling term can then be written as
\begin{equation}
	\label{eq:coupling}
	J \Delta  \sum_{\eta = 1}^3 \sum_{m,\, \mathbf k} \sum_{\alpha,\beta}  (\hat{\mathbf n}_{\eta}\cdot \bm\sigma_{\alpha\beta}) \zeta^m_{\eta}
	\, w^{m\,*}_{\mathbf k} w^m_{\mathbf k + \mathbf Q_\eta}  f^\dagger_{\mathbf k, \alpha} f^{\;}_{\mathbf k + \mathbf Q_{\eta}, \beta} \;.
\end{equation}
The dominant contributions come from states at the three saddle points, i.e. for electrons with $\mathbf k \approx \mathbf Q_\eta$.  At the three saddle points for filling fraction $f = 1/4$, we have $\vec{ w}_{\mathbf Q_1} = (0, 1, 1)$, $\vec{ w}_{\mathbf Q_2} = (1, 0, 1)$, and $\vec{ w}_{\mathbf Q_3} = (1, 1, 0)$.  Consequently, the product $w^m_{\mathbf Q_\mu} w^m_{\mathbf Q_\nu} = |\varepsilon_{m,\mu,\nu}| $ (no summation over $m$), where $\varepsilon$ is the antisymmetric tensor of rank 3. This result immediately implies that the coupling will be maximized when $\zeta^m_{\eta} = \delta_{m,\eta}$, i.e. it has the  $A_1$ irreducible representation of $C_{6v}$ discussed above. The corresponding magnetic order thus transforms according to the $A_2$ irreducible representation, consistent with our numerical result.

Interestingly, we find numerically that the same chiral magnetic order is also the ground state at filling fraction $f = 7/12$. This is consistent with the observation that a gap is opened at this filling fraction for strong enough coupling [Fig.~\ref{fig:vH1}(b)]. Snapshots of the spin configuration obtained from our large-scale KPM-Langevin simulations are shown in Fig.~\ref{fig:snapshot1} for the two filling fractions. The coloring of triangles indicates the amplitude of the scalar chirality $\kappa_{\triangle} = \mathbf S_1\cdot\mathbf S_2 \times \mathbf S_3$. Both snapshots clearly show large domains of uniform chirality but, curiously, with  different defect structures, indicating subtle differences in the effective spin-spin interactions.

\subsection{Van Hove filling $f = 5/12$~\label{filling2}}

\begin{figure}
\includegraphics[width=0.99\columnwidth]{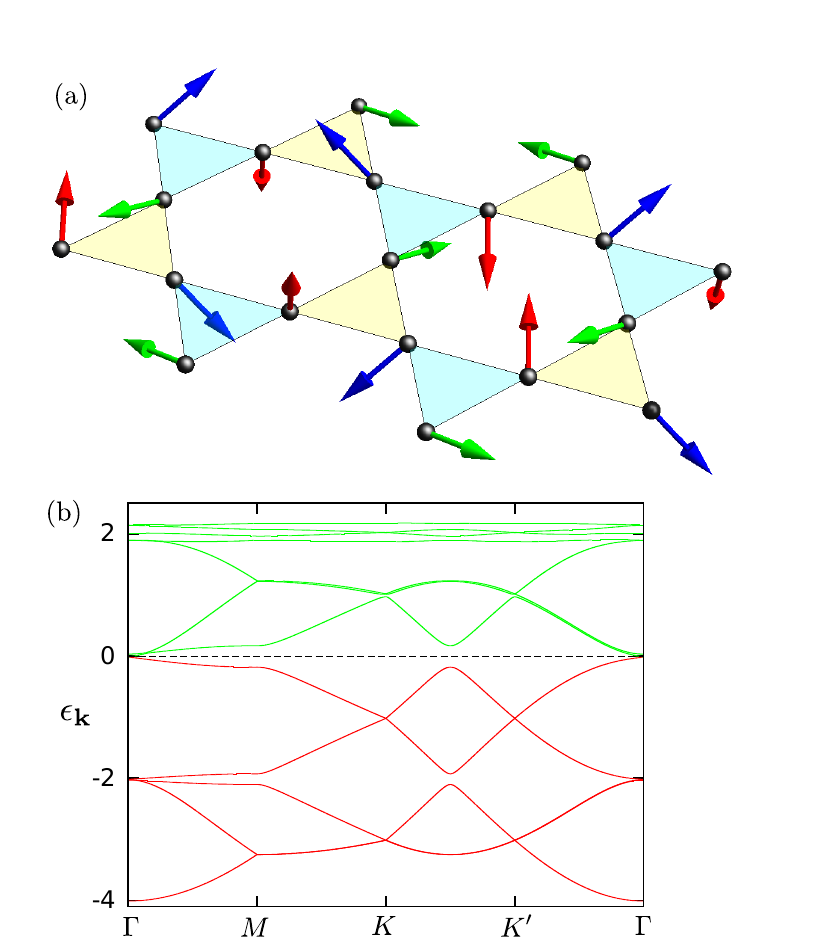}
\caption{\label{fig:vH2} (a) The $B_2$ symmetry triple-$\mathbf Q$ noncoplanar spin order on kagome lattice. Spins at the three inequivalent sublattices are indicated by different colors.  This state is the ground state at filling fraction $f = 5/12$. Panel (b) shows the corresponding band structure calculated with exchange coupling $J = 0.2\,t$. The dashed line indicates the Fermi level at $f = 5/12$ filling.
}
\end{figure}

\begin{figure*}
\includegraphics[width=1.98\columnwidth]{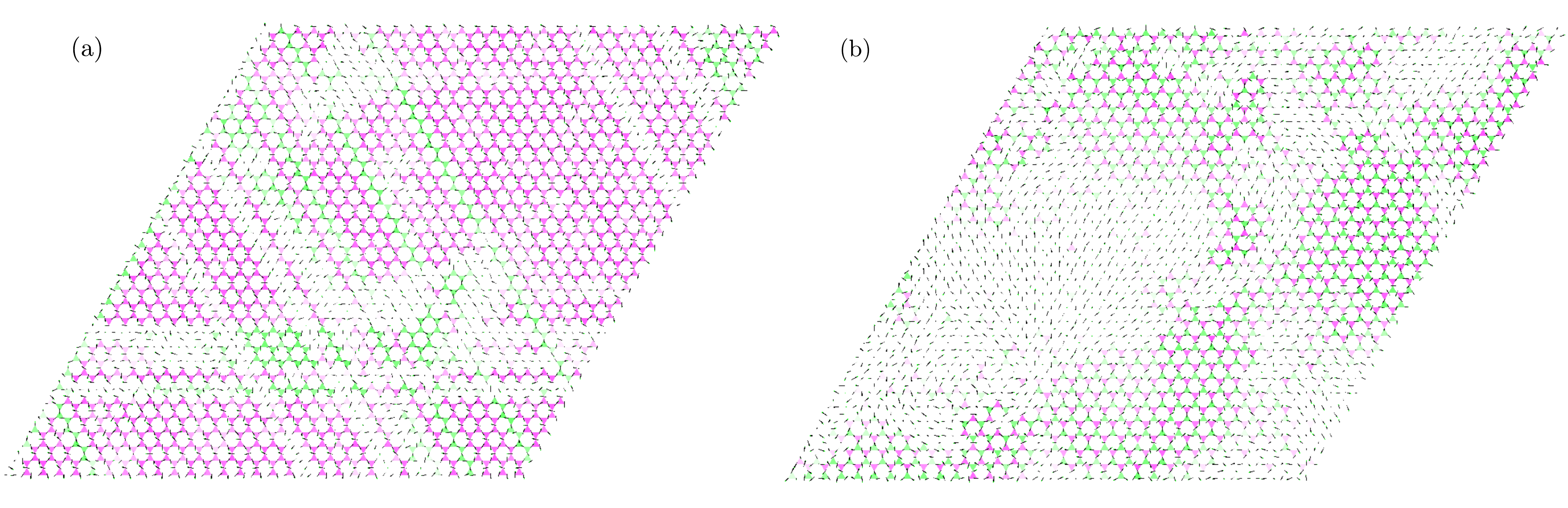}
\caption{\label{fig:snapshot2} Snapshots of two metastable spin configurations obtained from large-scale KPM-Langevin simulations for filling fraction $f=5/12$. The color indicates the $z$-component of the vector chirality $\bm\chi_{\triangle} =\mathbf S_1\times\mathbf S_2 + \mathbf S_2 \times \mathbf S_3 + \mathbf S_3 \times \mathbf S_1$. The two panels show (a) ferromagnetic and (b) antiferromagnetic alignments of $\bm\chi_{\triangle}$. These two $120^\circ$ coplanar orderings have nearly identical energy densities, so entropy and non-equilibrium effects determine the ordering found in our KPM-Langevin simulations. 
}
\end{figure*}

The numerically observed magnetic ordering at the second Van Hove singularity is shown in Fig.~\ref{fig:vH2} along with the corresponding dispersions. This $B_2$ symmetry spin state is also confirmed by the variational Fourier-space calculation. Again we find a vanishing uniform component $\bm\Delta_0 = \mathbf 0$. The non-zero order parameters at the nesting wavevectors are:
\begin{eqnarray}
	\label{eq:B2-state}
	&  & \bm\Delta^{1}_1 = +(S/\sqrt{2})\,\hat{\mathbf n}_1 \;, \quad \bm\Delta^{3}_1 = -(S/\sqrt{2})\,\hat{\mathbf n}_1 \;, \nonumber \\
		&  & \bm\Delta^{1}_2 = +(S/\sqrt{2})\,\hat{\mathbf n}_2 \;, \quad \bm\Delta^{2}_2 = -(S/\sqrt{2})\,\hat{\mathbf n}_2 \;, \quad  \\
			&  & \bm\Delta^{2}_3 = +(S/\sqrt{2})\,\hat{\mathbf n}_3 \;, \quad \bm\Delta^{3}_3 = -(S/\sqrt{2})\,\hat{\mathbf n}_3 \;. \nonumber 
\end{eqnarray}
Here \{$\hat{\mathbf n}_1$, $\hat{\mathbf n}_2$, $\hat{\mathbf n}_3$\} are again three orthogonal unit vectors. Each sublattice  participates in two of the nesting order parameters, while spins in each sublattice are coplanar. The normal of the coplanar spins from different sublattices are orthogonal to each other.  Similar to the previous case, the electronic bands are doubly degenerate. The $B_2$ symmetry state at $f = 5/12$ filling is rather robust for $J$ up to order $t$.

Interestingly, there remains a quadratic Fermi point at $\mathbf k = {\bf 0}$ for $f =5/12$ filling; see Fig.~\ref{fig:vH2}(b). The lack of the spectral gap at this Van Hove filling as compared to the state at $f = 1/4$ can also be understood from the couplings between the saddle points. As discussed in Sec.~\ref{sec:f1_4}, the effective coupling between electrons at the three inequivalent saddle points is proportional to $\zeta^m_\eta w^m_{\mathbf Q_\mu} w^m_{\mathbf Q_\nu}$ [Eq.~(\ref{eq:coupling})]. For filling fraction $f = 5/12$, the sublattice weights obtained by diagonalizing the hopping matrix in Eq.~(\ref{eq:hopping}) are $w^m_{\mathbf Q_\mu} = \delta_{m, \mu}$. Since the nesting wavevectors always connect different saddle points, the product $w^m_{\mathbf Q_\mu} w^m_{\mathbf Q_\nu} = 0$ for $\mu \neq \nu$ (no summation over $m$), hence the coupling coefficient vanishes  at the $M$-points. Consequently, the three saddle points remain degenerate, giving rise to a Fermi point at $\mathbf k = {\bf 0}$ in the reduced BZ. The vanishing couplings between the saddle points are related to the nontrivial sublattice interference discussed in Refs.~\cite{kiesel12,kiesel13}. Although the susceptibility at the three $\mathbf Q_\eta$ still has a logarithmic divergence (instead of $\log^2$) due to the divergent DOS at the saddle points, the triple-$\mathbf Q$ spin order now has to compete with simple ferromagnetism instability which also results from a divergent DOS. Indeed, the $\mathbf Q = {\bf 0}$ ferromagnetic order is the dominant SDW instability in Hubbard-like model on kagome lattice~\cite{wang13,kiesel13}. For Kondo-lattice model, our results show that the triple-$\mathbf Q$ spin order is still favorable energetically, and the selection of the $B_2$ symmetry state shown in Fig.~\ref{fig:vH2}(a) is due to higher-order couplings on the nested Fermi surface.

It is worth noting that spins in each individual triangle (both up and down) form a 120$^\circ$ coplanar structure in the $B_2$ symmetry spin order. This triple-$\mathbf Q$ state is thus one of the many classical ground states of the well-studied nearest-neighbor exchange interaction Hamiltonian $\mathcal{H}_{\rm NN} = J_{\rm AF}\sum_{\langle ij \rangle} \mathbf S_i\cdot\mathbf S_j$  with antiferromagnetic coupling constant $J_{\rm AF} > 0$~\cite{chalker92,chern13}.  Since the NN interactions can be recast into the form $\mathcal{H}_{\rm NN} = (J_{\rm AF}/2) \sum_{\triangle} |\mathbf M_{\triangle}|^2$ up to an irrelevant constant. A triangle with three spins pointing 120$^\circ$ to each other has a zero total spin $\mathbf M_{\triangle} = {\bf 0}$. Consequently, any spin configuration on kagome consisting of 120$^\circ$ triangles is a ground state of $\mathcal{H}_{\rm NN}$ and the $B_2$-symmetry state is one of them. In fact it is known that all such classical ground states with local $\mathbf M_{\triangle} = {\bf 0}$ form an extensively degenerate manifold~\cite{chalker92,chern13}.  Our large-scale KPM-Langevin simulations of the KLM at $f=5/12$ filling show that spins tend to form $120^\circ$ local order as temperature is lowered, indicating that the dominant effective spin interactions can be well-described by a NN exchange Hamiltonian. 

To characterize the general ground states of $\mathcal{H}_{\rm NN}$ with local $\mathbf M_{\triangle} = {\bf 0}$, it is useful to consider the configuration of the vector chirality $\bm\chi_{\triangle} =\mathbf S_1\times\mathbf S_2 + \mathbf S_2 \times \mathbf S_3 + \mathbf S_3 \times \mathbf S_1$ for individual triangles.  The noncoplanar $B_2$ symmetry spin state shown in Fig.~\ref{fig:vH2}(a) corresponds to a specific tetrahedral ordering of the the vector chirality $\bm\chi_{\triangle}$.  As shown in Fig.~\ref{fig:chiral}, the vector chirality at the four inequivalent triangles points to the four corners of a regular tetrahedron. The tetrahedral ordering of the vector chiralites thus results from the long-range (beyond NN) electron-mediated interactions in the KLM, selecting this particular spin configuration out of the degenerate set of ground states of $\mathcal{H}_{\rm NN}$.  Although this triple-$\mathbf Q$ state is the ground state of the KLM, hence is energetically favorable as temperature tends to zero, our large-scale KPM-Langevin dynamics simulations find other coplanar 120$^\circ$ states that are very close in energy. As shown in Fig.~\ref{fig:snapshot2}, the vector chiralities tend to colinearly align (parallel or antiparallel) at finite temperatures. These results are consistent with the fact that thermal fluctuations favor coplanar spins (order-by-disorder) in the NN exchange model~\cite{chalker92,chern13}.

The electronic subsystem must have zero Chern number because the scalar chirality is zero for coplanar spins in each triangle. Alternatively, the presence of reflection symmetries in the representation $B_2$, discussed in Section~\ref{symmetry}, forces the Chern number to be zero~\cite{fang12}.  However, the magnetic structure itself is chiral ($\kappa \neq 0$), as already discussed in Section~\ref{symmetry}. This is also manifested by the tetrahedral order of the vector chirality.

\section{Special Fermi points~\label{fermipoints}}

We now come to the fillings corresponding to the isolated band touching points of kagome tight-binding Hamiltonian. In this Section we investigate the magnetic ordering when the Fermi surface shrinks to these special points at filling fractions $f = 1/3$ and 2/3.  In the first case, the elementary excitations are dominated by electrons in the vicinity of two Dirac points. The Dirac points are connected by momentum vectors $\mathbf K_{\pm} = (\pm 4\pi/3, 0)$ and coupling between the Dirac point would therefore require magnetic ordering at these wavevectors. In contrast, at filling fraction $f = 2/3$, the Fermi surface shrinks to a single point at the zone center ($\mathbf k = \mathbf 0$), where the dispersive $\epsilon_2$ band touches the flat band. It is thus difficult to anticipate what the order parameter components will be other than the ${\bf q} =\mathbf 0 $ component. Surprisingly, our large-scale simulations uncover another triple-$\mathbf Q$ magnetic ordering, which has different symmetry from those induced at the Van Hove fillings. We first discuss the Dirac points and then come to the quadratic band crossing.

\subsection{Dirac points at $f = 1/3$~\label{dirac}}

\begin{figure}
\includegraphics[width=0.83\columnwidth]{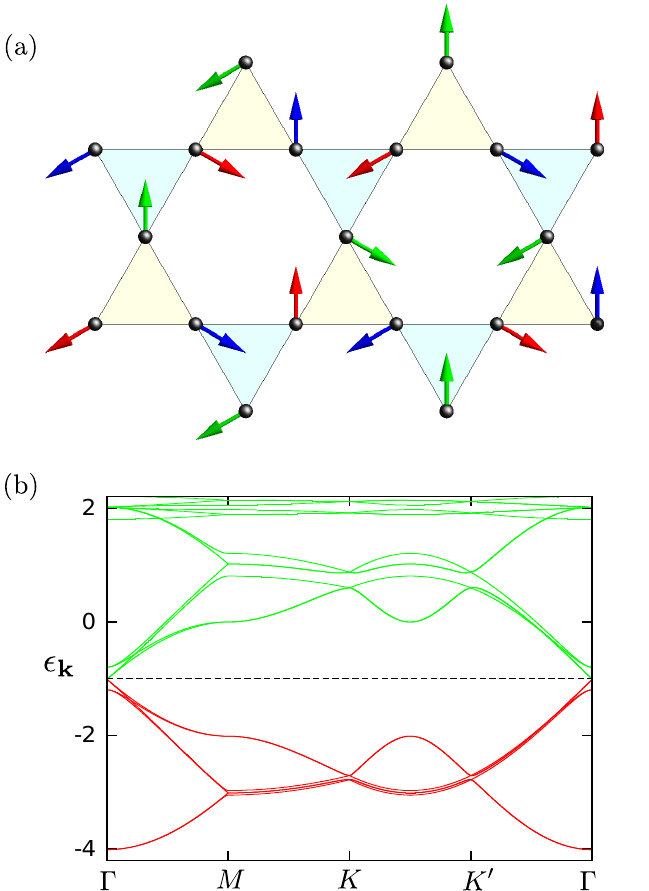}
\caption{\label{fig:dirac} (a) The $\sqrt{3}\times\sqrt{3}$ coplanar spin order on kagome lattice at filling fraction $f = 1/3$. Spins at the three inequivalent sublattices are indicated by different colors. (b) The corresponding band structure with exchange coupling $J = 0.2\,t$. The dashed line indicates filling fraction $f = 1/3$.
}
\end{figure}

The Fermi `surface' at filling fraction $f = 1/3$ consists of two Dirac points located at the coners of the Brillouin zone.  Elementary excitations are then dominated by electrons in the vicinity of these two points.  Although the two Fermi points are connected by wavevectors $\mathbf K_{\pm} = (\pm 4\pi/3, 0)$ (see Fig.~\ref{fig:dispersion}), the corresponding susceptibility is not divergent and it only shows a finite maximum, instead of a divergence, due to the vanishing DOS at the Dirac points. Indeed, the magnetic ordering obtained with our KPM-Langevin dynamics has the nesting wavevectors $\mathbf K_{\pm}$ shown in Fig.~\ref{fig:dirac}(a).  Interestingly, this spin configuration coincides with the famous coplanar $\sqrt{3}\times \sqrt{3}$ structure, which is also the ground state of the NN antiferromagnetic Heisenberg Hamiltonian $\mathcal{H}_{\rm NN}$  (spins in each triangle form a 120$^\circ$ structure with zero total moment). This suggest that NN exchange is the dominant electron-mediated spin-spin interaction, similar to the case of the second Van Hove singularity.  The $\sqrt{3}\times\sqrt{3}$ structure with staggered vector chiralities $\bm\chi_{\triangle}$ on the two inequivalent triangles is selected by longer-range terms of the effective interaction. Our KPM-Langevin simulations show that this coplanar state is stable up to $J \sim 0.5 t$.

A symmetry-based approach analogous to the one described in Sec.~\ref{symmetry} for the $M$-point ordering can be applied to the $K$-point case, i.e., the valleys at which the Dirac points are located for filling fraction $f=1/3$. The little group of the $\mathbf {K}_{\pm} = \pm \mathbf K = \pm (4\pi/3, 0) $ vectors is $C_{3v}$. In addition, $2 \mathbf {K}_{+} = \mathbf {K}_{-}$ modulo a reciprocal lattice vector. Following the same reasoning, we first find all scalar orders which transform as $1D$ irreducible representations of the kagome lattice symmetry group. We identify two distinct ordered states modulated by the valley momenta $\mathbf K$. By embedding them in a spinful setting, with the requirement of translational invariance modulo global rotations, we obtain a  single spin ordered state: the coplanar $\sqrt{3}\times \sqrt{3}$ state of Fig.~\ref{fig:dirac}. This result can be  intuitively understood by considering the two independent functions $\cos(\mathbf {K}\cdot \mathbf {r})$ and $\sin(\mathbf {K}\cdot \mathbf {r})$ defined on a triangular Bravais lattice. These two functions transform as $1D$ representations of $C_{6v}$. One can then embed them in a spinful  setting as $\mathbf {S}_m(\mathbf {r}) = \cos(\mathbf {K}\cdot \mathbf {r})\, \hat{\mathbf e}^m_1 + \sin(\mathbf {K}\cdot \mathbf {r})\, \hat{\mathbf e}^m_2$, where $m = 1,2,3$ denotes the three sublattices of kagome lattice. The three unit vectors $\hat{\mathbf e}^m_1$ form a 120$^\circ$ structure with zero vector sum, and $\hat{\mathbf e}^m_2 = \hat{\bm \chi}\times \hat{\mathbf e}^m_1$, where $\hat{\bm\chi}$ is the normal of the coplanar $\hat{\mathbf e}^1_1$ and $\hat{\mathbf e}^1_2$ vectors. 

Interestingly, the electron system remains gapless; the $\Gamma$-point of the reduced Brillouin zone remains doubly degenerate, in addition to the degeneracy required by the combined translation-rotation. This additional degeneracy can be understood by examining the interaction matrix between the two valley Dirac points. The vector order parameter with momenta $\mathbf K_{\pm}$ has the form $\bm\Delta^m_{\pm} = \Delta(\hat{\mathbf e}^m_1 \pm  i \hat{\mathbf e}^m_2)$. The exchange coupling in this case becomes
\begin{equation*}
	J \Delta \sum_m \sum_{\alpha,\beta} \bm\sigma_{\alpha\beta}\!\cdot\! (\hat{\mathbf e}^m_1 + i \hat{\mathbf e}^m_2) 
	w^m_{ K_1} w^m_{ K_2} f^\dagger_{K_1, \alpha} f^{\;}_{K_2, \beta} + {\rm h.c.} \;,  
\end{equation*}
where the eigenvectors  at the two Dirac points are  $\vec{ w}_{ K_1} = (-1, 0, 1)$ and $\vec{ w}_{ K_2} = (0, 1, 1)$. Substituting these eigenvectors into the above expression, we find that the  interaction matrix in the valley-spin space has the form $\Gamma_{a\alpha, b\beta} = J \Delta \bigl[\tau^x_{ab} (\sigma^x_{\alpha\beta} + i \sigma^y_{\alpha\beta}) + {\rm h.c.} \bigr]$ where $\tau$ is Pauli matrices acting on the valley space and $a, b = 1,2$ is the valley index.  Straightforward diagonalization of this $4\times 4$ matrix gives a double degenerate eigenvalue at $\epsilon = 0$ and two nonzero $\epsilon = \pm 2 J \Delta$, consistent with the numerical calculation of the electron band structure for the $\sqrt{3}\times\sqrt{3}$ order [Fig.~\ref{fig:dirac}(b)] .

\subsection{Quadratic Fermi point at $f = 2/3$~\label{qbc}}

\begin{figure}[t]
\includegraphics[width=0.94\columnwidth]{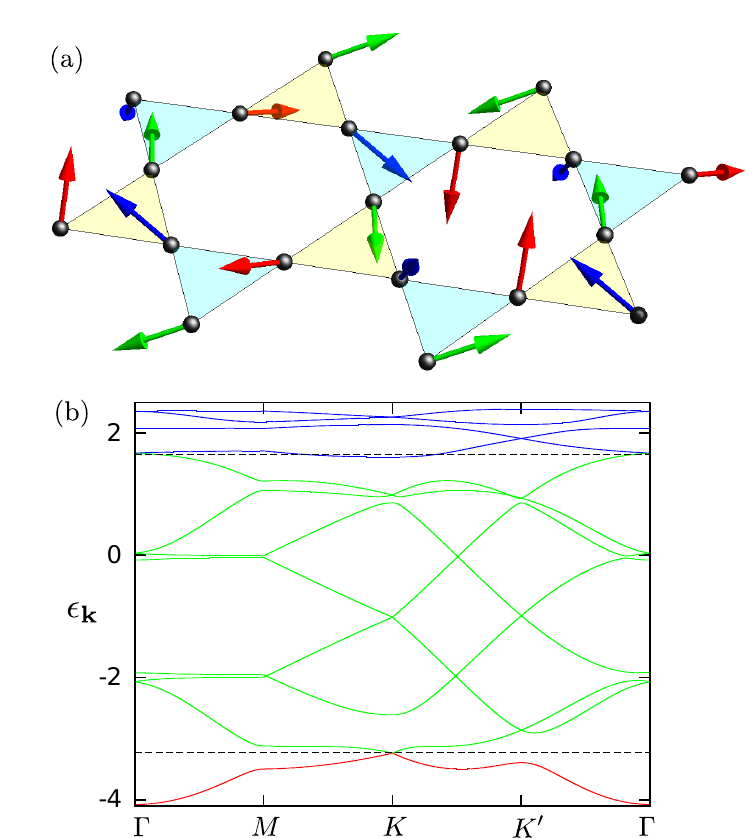}
\caption{\label{fig:qfp}  (a) The $B_1$ symmetry triple-$\mathbf Q$ noncoplanar spin order on the kagome lattice. Spins at the three inequivalent sublattices are indicated by different colors. It is the ground state at filling fraction $f = 2/3$ for weak couplings. The same magnetic order is also the ground state at filling fraction $f = 1/12$ for intermediate couplings. (b) The corresponding band structure with exchange coupling $J = 0.4\, t$. The two dashed lines indicate the Fermi levels at  $f = 1/12$ and $f = 2/3$ filling fractions.
}
\end{figure}

The magnetic ordering of the KLM at filling fraction $f = 2/3$ represents a difficult degenerate perturbation problem. The dispersive $\epsilon_2$ band touches the flat band $\epsilon_3$ at the $\Gamma$-point.  Consequently, the two $|\mathbf k = \mathbf 0\rangle$ states (with $\epsilon_{\mathbf k = \mathbf 0} = +2t$) from the dispersive band can couple to the extensively degenerate  states with nonzero momentum in the flat band.  Our KPM-Langevin dynamics simulations show that the ground state at filling fraction $f = 2/3$ is another triple-$\mathbf Q$ magnetic order, shown in Fig.~\ref{fig:qfp}(a), that transforms according to the $B_1$ irreducible representation discussed above. This state is characterized by the following non-zero order parameters
\begin{eqnarray}
	\label{eq:B1-state}
	&  & \bm\Delta^{1}_1 = +(S/\sqrt{2})\,\hat{\mathbf n}_1 \;, \quad \bm\Delta^{3}_1 = +(S/\sqrt{2})\,\hat{\mathbf n}_1  \;, \nonumber \\
		&  & \bm\Delta^{1}_2 = +(S/\sqrt{2})\,\hat{\mathbf n}_2 \;, \quad \bm\Delta^{2}_2 = +(S/\sqrt{2})\,\hat{\mathbf n}_2 \;, \quad  \\
			&  & \bm\Delta^{2}_3 = +(S/\sqrt{2})\,\hat{\mathbf n}_3 \;, \quad \bm\Delta^{3}_3 = +(S/\sqrt{2})\,\hat{\mathbf n}_3 \;, \nonumber 
\end{eqnarray}
where $\{\hat{\mathbf n}_1, \hat{\mathbf n}_2, \hat{\mathbf n}_3\}$ are three orthogonal unit vectors. The magnetic moments are coplanar within each sublattice, similar to the state shown in Fig.~\ref{fig:vH2}(a). The presence of reflection symmetries in this magnetic order again precludes topologically nontrivial electronic states~\cite{fang12}. This is consistent with the fact that the system remains gapless [Fig.~\ref{fig:qfp}(b)] at the $\Gamma$-point~\cite{sun09,chern12}. 

\begin{table*}
\centering
\begin{tabular}{l   l   l   l}
\hline
Filling fraction & Fermi surface singularity & Magnetic order & Electronic state \\
\hline
\hline
$f=1/12$ & No singularity & Noncoplanar $B_1$ symmetry order & Dirac semi-metal \\
\hline
$f=3/12$ & Van Hove singularity & Noncoplanar $A_2$ symmetry order & Quantum Hall insulator \\
\hline
$f=4/12$ & Dirac Fermi point & Coplanar $\sqrt{3}\times\sqrt{3}$ order & Dirac semi-metal \\
\hline
$f=5/12$ & Van Hove singularity & Noncoplanar $B_2$ symmetry order & Quadratic Fermi-point semi-metal \\
\hline
$f=7/12$ & No singularity & Noncoplanar $A_2$ symmetry order & Quantum Hall insulator \\
\hline
$f=8/12$ ($t_2 = 0$) & Quadratic Fermi point & Noncoplanar $B_1$ symmetry order & Finite Fermi surfaces (electron + hole pockets) \\
\hline
$f=8/12$ ($t_2 \neq 0$) & Quadratic Fermi point & Noncoplanar umbrella order & Quantum Hall insulator \\
\hline
\end{tabular}
\caption{\label{tab-summary} Summary of the ground-state magnetic orders and the corresponding electronic states at various commensurate filling fractions.}
\end{table*}

To shed light on the selection of the triple-$\mathbf Q$ order, we note that the couplings between electrons are proportional to the product $w^m_{\mathbf k} w^m_{\mathbf k'}$ [Eq.~(\ref{eq:coupling})].  At filling fraction $f = 2/3$ the basis of the doubly degenerate $\mathbf k = \mathbf 0$ states are $\vec{w}_{\Gamma} = (1, 0, -1)$ and $(1, -1, 0)$. Any $\mathbf k = \mathbf 0$ state with eigenenergy $\epsilon = +2t$ can be expressed as a linear combination of these two eigenvectors.  The eigenvectors at the three $M$-points with energy $\epsilon_M = 2 t$ are $\vec{ w}_{M_1} = (1, 0, -1)$, $\vec{w}_{M_2} = (0, 1, -1)$ and $\vec{ w}_{M_3} = (1, -1, 0)$, which lie completely within this subspace and thus give rise to the largest overlap with the $\mathbf k = \mathbf 0$ states.

Another interesting spectral feature of the $B_1$ symmetry triple-$\mathbf Q$ state is the appearance of Dirac nodes at only {\it one} of the two inequivalent $K$-points [Fig.~\ref{fig:qfp}(b)]. Since the two $K$-points are related by time-reversal symmetry, which valley point remains gapless depends on the sign of the magnetic order parameter. The fact that the Dirac nodes should appear only at one of the two valleys is again in agreement with the symmetry-breaking of the $B_1$ irreducible representation. For large enough couplings, the remaining Dirac point is an isolated crossing between the lowest two bands, i.e. there is no overlapping between the two bands. With the aid of unbiased KPM-Langevin dynamics simulations, we find that the magnetic order shown in Fig.~\ref{fig:qfp}(a) is also the ground state at filling fraction $f = 1/12$ for intermediate coupling constants $J \gtrsim 0.3 t$. The Fermi ``surface'' thus shrinks to the remaining Dirac nodes at one of the $K$-points. Further symmetry-breaking perturbation can gap out the residual Dirac point and give rise to a topological insulating state.  

The dispersion of the $\epsilon_2$ band in the vicinity of this Fermi point is quadratic. The finite DOS at such a Fermi point renders it susceptible to weak perturbations~\cite{sun09}. Moreover, the $2\pi$ Berry flux around this band-crossing point indicates that gapping out the Fermi point would lead to a topologically nontrivial insulating state. However, the existence of a flat band is a rather special limit in real systems. Indeed, inclusion of, e.g. next-nearest-neighbor hopping $t_2$ lifts the flat-band degeneracy. It is thus also interesting to investigate the magnetic ordering of the KLM at the same $f = 2/3$ filling in the presence of a small $t_2$. To avoid unnecessary complications, we consider a small {\em negative} $t_2/t$, such that the new $\epsilon_3$ band is bending upward, i.e. the minimum of $\epsilon_3$ is at the $\Gamma$-point. In this situation, we are left with an isolated quadratic band crossing point at $\mathbf k = \mathbf 0$. The magnetic ground state is expected to have a $\mathbf Q = \mathbf 0$ long-range order. By performing unbiased large-scale KPM-Langevin dynamics simulations at, e.g. $t_2 = - 0.1 t$ and $J = 0.1 t$, we indeed find a $\mathbf Q = \mathbf 0$ magnetic order, in which spins form a nearly 120$^\circ$ umbrella structure in each triangle~\cite{ohgushi00}. Although the out-of-plane canting angle is rather small ($ < 10^\circ$), the quadratic Fermi point is gapped out by the spin noncoplanarity and the electronic ground state is a QAH insulator~\cite{Taillefumier06}; see Appendix~\ref{appendix2} for more details.

\section{Conclusions~\label{conclusions}}

The band structure of the kagome lattice gives rise to a rich variety of singular Fermi surfaces at certain commensurate filling fractions. Therefore, in this work we have addressed the question: What  weak-coupling instabilities of the KLM  are triggered by these singularities? We have presented a systematic symmetry-based approach has been shown to considerably limit the number of candidate spin ground states. The full variational space of potential magnetic orderings is determined by identifying the  wave-vectors that maximize the magnetic susceptibility of the conduction electrons. This space is not necessarily small for lattices, such as kagome, that have more than one atom per unit cell. The potential magnetic orderings are classified according to the irreducible representations of the small group of the ordering vectors. We have demonstrated that imposing symmetry constraints reduces the large variational space to a small number of candidate ground states. Our main results are summarized in Table~\ref{tab-summary}.

To determine the ground state magnetic ordering of the kagome KLM for specific fillings, we have complemented our analytical study with two complementary numerical techniques. First, we have performed a $T = 0$ variational Fourier-space calculation, assuming a certain magnetic unit cell. Second, we have performed large-scale numerical simulations based on the recently developed KPM-Langevin algorithm. This method makes it practical to simulate the very large lattice sizes required to capture logarithmic divergences in the magnetic susceptibility. Our numerical results coincide with the results of our analytical treatment when the latter is applicable, and prove that the high-symmetry spin configurations are  good variational states. 

The kagome lattice was shown to have three distinct spectral features, giving rise to very interesting spin textures for the corresponding electron fillings. At the two Van Hove fillings, i.e. $f=1/4$ and $f=5/12$, we have found spin states modulated by the three inequivalent nesting vectors, both noncoplanar and chiral. While the corresponding electron state at $f = 1/4$ is quantum Hall insulator, the other one preserves a Fermi point at the $5/12$ filling. The filling fractions $f=1/3$ and $f=2/3$ exhibit Dirac points and a Quadratic Band Crossing, respectively. At the Dirac point filling we find the well-known coplanar $\sqrt{3}\times \sqrt{3}$ magnetic ordering. The triple-${\bf Q}$ magnetic ordering obtained numerically for $f=2/3$ (which is different from the triple-${\bf Q}$ ordering observed at the Van Hove fillings) is quite unexpected, if we naively assume that the magnetic ordering is triggered by the instability of a quadratic Fermi point, which is located at the zone center. Indeed, by lifting the full degeneracy of the flat-band with a finite second-nearest-neighbor hopping $t_2$, we find a $\mathbf Q = 0$ magnetic order, as expected from an isolated quadratic band crossing point. In addition to these special filling fractions, we have considered two other fillings, which are commensurate with a quadrupling of the magnetic unit cell implied by triple-${\bf Q}$ ordering. They correspond to $f=1/12$ and $f=7/12$. Intriguingly, at $f=7/12$ we find the same triple-${\bf Q}$ state as for $f=1/4$. The electronic spectrum is gapped for large enough $J$. At $f=1/12$ we find a magnetic ground state which leads to isolated degeneracies, i.e. Dirac points.

Spontaneous Chern insulators appear for different filling factors of the kagome lattice, i.e. $f=1/4$, $f=7/12$, and $f=2/3$. By considering that similar results have been obtained for other lattices~\cite{martin08,kato10,Akagi10,Venderbos11,Venderbos12,Hayami14,chern10,Ishizuka13,Ishizuka13b}, we can safely conclude that Kondo-lattice systems are rather strong candidates for realizing spontaneous  QAH states at ambient temperature. While realistic band structures are much more complex  than the simple tight-binding models considered here, and in previous works,   the weak-coupling instabilities that we have discussed here are only sensitive to singular features of the Fermi surface. Because these singular features also appear in more realistic band structures, our main conclusions remain relevant. 

We also note that noncoplanar magnetic ordering is a prerequisite for inducing the QAH effect. The converse, however, is not true. Not all noncoplanar spin orderings lead to a QAH effect, as we have shown based on symmetry arguments. Importantly, our results confirm that, in contrast to local moment magnets, noncoplanar orderings are quite ubiquitous in itinerant magnets. 

Finally, the rather large lattice sizes amenable to the KPM-Langevin dynamics method also allow us to investigate magnetic ordering at arbitrary filling fractions and coupling strength. Since spin configurations at these conditions most likely are characterized by incommensurate wavevectors, large-scale unbiased numerical minimization (which is insensitive to boundary conditions) is required. Indeed, we have found unusual incommensurate structures and skyrmion arrays in the kagome KLM with intermediate coupling constant. A detailed characterization of these magnetic orders will be left for future studies. 

\appendix

\section{Variational Fourier-space minimization}
\label{appendix1}

In this Appendix, we outline the variational minimization method for magnetic orderings at the two Van Hove filling fractions. As discussed in the main text, the combination of divergent DOS at the saddle points and perfect Fermi surface nesting, Fig.~\ref{fig:dispersion}(c), gives rise to a logarithmically squared divergent susceptibility at the three nesting vectors $\mathbf Q_\eta$. One thus expects magnetic textures dominated exclusively by these three wavevectors. In real space, the most general spin state modulated by the three nesting vectors consist of a quadrupled unit cell with 12 spins. The four inequivalent up triangles of the kagome lattice are located at $\mathbf r + \mathbf a_k$, where $\mathbf r = 2l \mathbf a_1 + 2n \mathbf a_2$, ($k = 0, 1, 2, 3$, with $l,n$ integers) and we have introduced $\mathbf a_0 = \mathbf 0$ for simplicity. In terms of the order parameters, the different spins are expressed as $(m = 1,2,3)$
\begin{eqnarray}
	\label{eq:quadrupled}
	\mathbf S_m(\mathbf r + \mathbf a_0) &=& \bm\Delta^m_{0} + \bm\Delta^m_{1} + \bm\Delta^m_{2} + \bm\Delta^m_{3}\;, \nonumber \\
	\mathbf S_m(\mathbf r + \mathbf a_1) &=& \bm\Delta^m_{0} + \bm\Delta^m_{1} - \bm\Delta^m_{2} - \bm\Delta^m_{3}\;, \nonumber \\
	\mathbf S_m(\mathbf r + \mathbf a_2) &=& \bm\Delta^m_{0} - \bm\Delta^m_{1} + \bm\Delta^m_{2} - \bm\Delta^m_{3}\;, \nonumber \\
	\mathbf S_m(\mathbf r + \mathbf a_3) &=& \bm\Delta^m_{0} - \bm\Delta^m_{1} - \bm\Delta^m_{2} + \bm\Delta^m_{3}\;.
\end{eqnarray}
For completeness, we have included the $\mathbf q = \mathbf 0$ component $\bm\Delta^m_{0}$ corresponding to uniform ordering at each individual sublattice. Instead of working with the order parameters, we consider the most general spin configurations $\{\mathbf S_i \}$ ($i = 1, \cdots, 12$) within the extended unit cell. This leaves us with 24 independent variational parameters $\{\theta_i, \phi_i\}$ characterizing the orientations of classical spins with fixed length.

\begin{figure}[t]
\includegraphics[width=0.8\columnwidth]{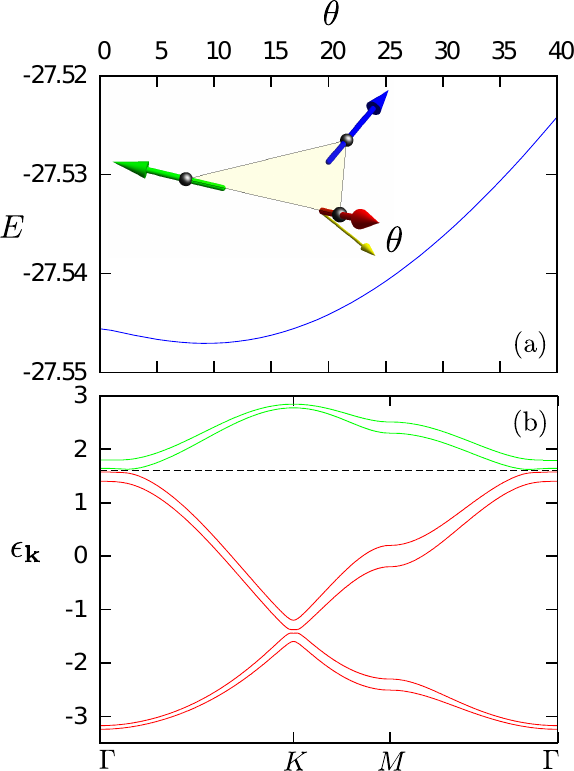}
\caption{\label{fig:umbrella}  (a) The total energy (arbitrary units) as a function of the canting angle $\theta$ for the $\mathbf Q = 0$ umbrella state. The inset shows the three spins of the umbrella state in a triangular unit cell. This curve is obtained using the following parameters: $J = 0.2 t$, $t_2 = -0.2 t$. (b) The corresponding band structure with exchange coupling $J = 0.2\, t$, $t_2 = -0.2 t$ and canting angle $\theta = 10^\circ$. The dashed line indicates the Fermi level at $f = 2/3$ filling.
}
\end{figure}

With the periodic spin structure specified in Eq.~(\ref{eq:quadrupled}), the electron Hamiltonian can be diagonalized using Fourier transformation. It contains two terms:
\begin{eqnarray}
	\mathcal{H} = \sum_{\mathbf k} \sum_{m,n=1}^3 \sum_{\eta,\xi=0}^3 \sum_{\alpha\beta} c^\dagger_{m\alpha}(\mathbf k + \mathbf Q_\eta) c^{\;}_{n\beta}(\mathbf k + \mathbf Q_\xi) \nonumber \\
	\quad\quad\times \Bigl[h_{mn}(\mathbf k+\mathbf Q_{\eta}) \,\delta_{\alpha\beta} \,\delta_{\eta\xi} + \mathcal{M}^{(m)}_{\eta\alpha;\,\xi\beta}\, \delta_{mn} \Bigr].
\end{eqnarray}
The $\mathbf k$ summation is over the reduced BZ.  The matrix $h_{mn}$ in the first hopping term is given in Eq.~(\ref{eq:hopping}), while the exchange coupling term is
\begin{eqnarray}
	\mathcal{M}^{(m)}_{\eta\alpha;\,\xi\beta}(\mathbf k) = -J \Theta_{\eta\xi;\,\zeta} (\bm\Delta^m_\zeta \cdot \bm\sigma_{\alpha\beta}) e^{i (\mathbf Q_\eta - \mathbf Q_\xi)\cdot \mathbf d_m},
\end{eqnarray}
where $\mathbf d_m$ denotes the basis vectors for the three sublattices. The factor $\Theta_{\eta\xi;\zeta}$ encodes the momentum conservation; it is symmetric with respect to the first two indices. $\Theta_{\eta\xi;\zeta} = 1$ when the three indices are all different and is zero otherwise. For a given spin state, the total energy at $T = 0$ is given by $E_0(\{\theta_i,\phi_i \}) = \sum_r \sum_{\mathbf k} \epsilon_{r, \mathbf k} \theta(\epsilon_F - \epsilon_{r, \mathbf k})$, where the eigenenergies $\epsilon_{r,\mathbf k}$ are obtained by diagonalizing the $24\times 24$ matrix: $H_{m\eta\alpha;\,n\xi\beta} = h_{mn}(\mathbf k + \mathbf Q_\eta)\delta_{\alpha\beta}\delta_{\eta\xi} + \mathcal{M}^{(m)}_{\eta\alpha;\,\xi\beta}\delta_{mn}$. We then employ the simulated annealing algorithm to minimize $E_0$ with respect to the angle parameters. Starting from random initial configurations and using different rates of decreasing the effective temperatures, we robustly obtain the two triple-$\mathbf Q$ states ($A_2$ and $B_2$ irreducible representations) at the respective Van Hove filling fractions.

\section{Umbrella state at $f=2/3$ filling}
\label{appendix2}

Here we consider the kagome Kondo-lattice model with an additional second-nearest-neighbor hopping $t_2 < 0$ at filling fraction $f = 2/3$. As discussed in the main text, the flat band degeneracy is lifted by the additional $t_2$, and an isolated quadratic band-crossing is left at the $\Gamma$ point. Our large-scale Langevin dynamics simulations find a $\mathbf Q = {\bf 0}$ state with spins in each triangle pointing almost at $120^\circ$ to one another. It is possible that the spins develop a uniform out-of-plane canting since such noncoplanar structure might completely gap out the quadratic Fermi point at $f = 2/3$ filling. In order to examine the ground-state spin order in more detail, we assume a $\mathbf Q = {\bf 0}$ umbrella structure with a uniform canting $\theta$ and compute the $T = 0$ total energy as a function of the canting angle; the result is shown in Fig.~\ref{fig:umbrella}(a) for $J = 0.2 t$ and $t_2 = -0.2 t$. Interestingly, the energy minimum is reached at a nonzero small canting angle $\theta \approx 10^\circ$. In general the canting angle is rather small; it depends on both the exchange coupling $J$ and the second-neighbor hopping $t_2$. Fig.~\ref{fig:umbrella}(b) shows the electronic band structure of the umbrella state. Both the Dirac and quadratic band crossing points are gapped out in this spin state. At $f = 2/3$ filling, the system is an insulator with spontaneous quantum Hall effect as pointed out in Ref.~\cite{Taillefumier06}.

\bigskip

{\it Note added.} -- Upon conclusion of this work, we became aware of overlapping results of a study on the same model~\cite{shivam-kagome}.

\begin{acknowledgements}
We thank useful discussions with S. Ghosh, M. Lawler, Y. Motome, and M. Udagawa. Work at LANL was carried out under the auspices of the U.S. DOE contract No.~DE-AC52-06NA25396  through the LDRD program. J.W.F.V acknowledges support from the Dutch Science Foundation NWO, and from FOM. The large-scale numerical simulations were performed using the CCS-7 Darwin cluster at LANL.
\end{acknowledgements}

\end{document}